\documentclass[useAMS,usenatbib]{mn2e}

\title[Stellar feedback and the ISM]
{The properties of the ISM in disc galaxies with stellar feedback}
\author[C. L. Dobbs]
{C. L. Dobbs\thanks{E-mail:
cdobbs@mpe.mpg.de}$^{1,2}$, A. Burkert$^{2,3}$ and J. E. Pringle$^{4}$\\
$^1$ Max-Planck-Institut f\"ur extraterrestrische Physik, Giessenbachstra\ss{}e, D-85748 Garching, Germany \\
$^2$ Universitats-Sternwarte M\"unchen, Scheinerstra\ss{}e 1, D-81679
M\"unchen, Germany \\
$^3$ Max-Plack fellow, Max-Planck-Institut f\"ur extraterrestrische Physik, Giessenbachstra\ss{}e, D-85748 Garching, Germany \\
$^4$ Institute of Astronomy, Madingley Road, Cambridge, CB3 0HA}
\usepackage{amssymb}
\usepackage{amsmath}
\usepackage{graphicx}
\usepackage{epsfig}
\usepackage{multirow}

\begin{document}
\date{\today}

\pagerange{\pageref{firstpage}--\pageref{lastpage}} \pubyear{0000}

\maketitle

\label{firstpage}
\begin{abstract}
We perform calculations of isolated disc galaxies to investigate how
the properties of the ISM, the nature of molecular clouds, and the
global star formation rate depend on the level of stellar feedback. 
We adopt a simple physical model, which includes a galactic
potential, a standard cooling and heating prescription of the ISM, and
self gravity of the gas.
Stellar feedback is implemented by
injecting energy into dense, gravitationally collapsing gas, 
but is independent of the Schmidt-Kennicutt relation.
We obtain fractions of gas, and filling
factors for different phases of the ISM in reasonable ageement with
observations. 
Supernovae are found to be vital to reproduce the scale
heights of the different components of the ISM, and velocity
dispersions. 
The GMCs formed in the simulations display mass spectra similar to the
observations, their normalisation dependent on the level of feedback. 
We find $\sim$40 per cent of the clouds exhibit
retrograde rotation, induced by cloud--cloud collisions.
The star formation rates we obtain are in good agreement with the
observed Schmidt-Kennicutt relation, and are not strongly dependent on the star
formation efficiency we assume, being largely self regulated by the feedback.
We also investigate the effect of spiral structure by comparing
calculations with and without the spiral component of the potential.
The main difference with a spiral potential is that more
massive GMCs are able to accumulate in the spiral arms. Thus we are able to
reproduce massive GMCs, and the spurs seen in many grand design
galaxies, even with stellar feedback. The presence of the
spiral potential does not have an explicit effect on the star formation
rate, but can increase the star formation rate indirectly by enabling the formation of
long-lived, strongly bound clouds.
\end{abstract}

\section{Introduction}
Star formation in galaxies is intrinsically linked to the evolution
and dynamics of the interstellar medium and in particular, molecular
clouds. Thus providing a plausible model of the evolution of galaxies,
and star formation within them requires being able to match the
properties of the ISM and molecular clouds, as well as reproducing a
star formation rate in agreement with the Schmidt-Kennicutt \citep{Schmidt1959,Kennicutt1998} relation.

In a previous paper \citep{Dobbs2011}, we examined one particular
aspect of molecular clouds. We demonstrated that with a simple model involving a standard cooling
prescription, UV heating and feedback from star formation, it was possible to
produce clouds with a distribution of virial parameters close to that
observed, and that through cloud-cloud collisions, and supernovae
feedback, the majority of the clouds were unbound. In the current paper we
consider many more aspects of our models, including the global
properties of the ISM, star formation rates, and the importance of
spiral structure, with an overall aim to match the observed
properties of galaxies, as described below.

The ISM in galaxies is characterised by a multiphase turbulent medium, regulated
by self gravity, cloud collisions, supernovae and stellar winds, galactic rotation,
spiral shocks and magnetic fields. These processes lead
to a velocity dispersion of 5--10~km~s$^{-1}$
\citep{derKruit1982,Dickey1990,Dickey1990a,Combes1997,vanZee1999,Petric2007,
Tamburro2009,Wilson2010} both in
HI and CO, in fact in all but the very hot gas. The main source of this
velocity dispersion however is not fully clear, although the recent consensus is that
turbulence is driven on large scales, e.g. by supernovae
or spiral shocks \citep{Ossenkopf2002,Brunt2003,Brunt2009,Padoan2009}. The thermal distribution of the ISM is not well
constrained but observations of the solar neighbourhood suggest that similar proportions of HI
lie in the cold, warm and intermediate regimes \citep{Heiles2003}. A potentially difficult
characteristic of the ISM to reproduce is the scale height of the
gas, particularly the cold phase. The presence of high latitute GMCs
such as Orion requires that molecular, and cold HI gas resides at distances
above the mid-plane much higher than expected simply from thermal
pressure. Previous simulations showed that in the
absence of stellar feedback, the scale height of the gas is too low, and
the distribution of HI in the $l-z$ plane does not match the
observations \citep{Douglas2010}.

In previous calculations, we showed that giant molecular clouds (GMCs)
form in spiral galaxies when smaller clouds are brought together in
the spiral arms \citep{Dobbs2008}. Self gravity aids this process
predominantly by increasing the frequency of collisions, and the
likelihood that clouds merge, due to the mutual gravitational
attraction of the clouds. An immediate question is whether GMCs can
still form by this means with stellar feedback, or whether feedback
disrupts smaller clouds before more massive GMCs can form \citep{Ostriker2007}.
In addition, the previous calculations without stellar feedback found cloud mass
functions in agreement with observations, and further showed that retrograde clouds are naturally
reproduced as a results of cloud collisions
\citep{Dobbs2008}. However these properties, at least the cloud mass spectrum,
are likely to be dependent on stellar feedback.

Observationally, the star formation rate is correlated with the global
surface density according to the Schmidt-Kennicutt relation. There is considerable scatter in
the star formation rate ($\gtrsim$ 1 order of magnitude) for a given
surface density, whilst the slope of the relation is found to vary
according to different gas tracers, from 1.0 for dense tracers such as CO
and HCN \citep{Wong2002,Gao2004,Bigiel2008,Genzel2010}, 
to 1.4 for the total gas surface density
\citep{Kennicutt1998,Wong2002} but with a steeper turnover below $\sim 10$
M$_{\odot}$ pc$^{-2}$. Nevertheless these observations provide our
best guide for linking galactic scale physics to small scale star
formation.

A fundamental question in relation to the properties of the ISM and
molecular clouds, and the star formation rates in galaxies is the
importance of spiral structure. \citet{Roberts1969} proposed that
spiral density waves could trigger star formation in the spiral arms. Thus
the presence of a spiral density wave
could explain some of the scatter observed in the Schmidt-Kennicutt relation.
However whether spiral shocks trigger star formation has been
strongly debated in the past. \citet{Elmegreen1986} examined star
formation rates in galaxies with different Hubble types, including
flocculent and grand design galaxies. Finding no variation in the star
formation rate for the different galaxy types, they concluded that
density wave triggering was only a small contribution to the star
formation rate. The main evidence of density wave triggered star
formation rates is a non-linear dependence of the star formation rate in
the spiral arms \citep{Cepa1990,Seigar2002}.  More recent
observations however again suggest that there is little difference in
the star formation efficiency
in the spiral arms \citep{Foyle2010}.
Thus the density wave may simply organise the gas and star
formation within a galaxy \citep{Tan2010}. In \citet{Dobbs2009} we
found that there was little dependence of the amount of bound gas on
the strength of the spiral potential, mainly because although the
density increases in the spiral shock, the velocity
dispersion also increases. In \citet{Dobbs2009}, the clouds are
also supported by magnetic fields.

Finally, whilst star formation occurs predominantly in the spiral arms,
many galaxies (typically those with well defined dust lanes) exhibit
star formation associated with interarm
spurs \citep{LaVigne2006}. Calculations without feedback
\citep{Kim2003,Wada2004, Shetty2006,Dobbs2007} show that spurs form
by the shearing out of spiral arm GMCs. However
it is unclear whether these structures would still form if the clouds
are disrupted by stellar feedback.

Simulations of isolated galaxies designed to model the ISM with
supernovae feedback were performed by \citet{Rosen1995}, and
later by \citet{Wada2000}. Both demonstrated that a 3-phase medium
is produced, and the higher resolution of the latter studies indicated
the complex structure of the ISM. However both were 2D, so unable to
calculate properties of molecular clouds. The difficulty of
performing high resolutions of a 3D disc has lead to many simulations
of kpc or so size periodic boxes of the ISM 
\citep{Korpi1999,deAvillez2000,deAvillez2004,deAvillez2005,Slyz2005}. These
show that supernovae feedback can reproduce the observed levels of turbulence
 \citep{Dib2006,Joung2006} and pressure \citep{Joung2009,Koyama2009} in the ISM. One other result
 from these calculations is that supernovae feedback appears to inhibit, rather
 than enhance, star formation \citep{Slyz2005,Joung2006}. These
 calculations are able to resolve turbulence and cooling
 instabilities on parsec scales. However they cannot examine processes such as molecular cloud
 formation by cloud collisions, or self gravity, which operate on larger
 scales.

Numerous calculations have modelled the ISM on galactic
scales with the aim of
reproducing the Schmidt-Kennicutt relation
\citep{Li2006,Tasker2008b,Robertson2008,Pelupessy2009,Gnedin2009}. 
\citet{Gnedin2009} also include a prescription
of molecular gas formation which they use to investigate different metallicity environments. 
One should note though that many calculations implicitly assume (or
explicitly in the case of cosmological simulations) some form of the 
Schmidt-Kennicutt relation \citep{Schmidt1959,Kennicutt1989} in adopting a star formation rate per unit
time. Other calculations explicitly insert the observed supernovae rate in
our Galaxy \citep{deAvillez2004,Shetty2008}.

A couple of calculations have also included a spiral potential and
stellar feedback in simulations of isolated discs
\citep{Shetty2008,Wada2008}. These tend to show that feedback is
fairly disruptive, in the case of \citet{Shetty2008} even destroying
the spiral structure, thus they have difficulty reproducing the spurs
observed in galaxies.

Our aim in this paper is to provide a
coherent, and global, description of the dynamics of the ISM, with
relatively simple physical processes and a minimum of assumptions. 
The paper is organised as follows.
We first describe how the
calculations are set up in Section 2. We then provide a generic picture
of the typical evolution of our calculations in Section 3. In Section~4 we present simulations
where we investigate the structure of the disc and properties of the
ISM for calculations with different star formation efficiencies, 
all with a spiral potential. In Section~5, we compare results
from calculations with and without a spiral potential. In Section~6,
we discuss results from higher surface density calculations, and
show the dependence of the global star formation rate on surface
density. We calculate the properties of GMCs from a selection of
 the calculations in Section~7. Finally in Section~8 we present our conclusions.   

\section{Calculations}
The calculations presented here are 3D SPH simulations using an SPH code
developed by Benz \citep{Benz1990}, Bate \citep{Bate1995} and Price
\citep{PM2007}. In all the calculations presented here, the gas is assumed to orbit in a fixed galactic
gravitational potential. The potential is a logarithmic potential
\citep{Binney1987}, which produces a flat rotation curve with a maximum velocity of 220 km~s$^{-1}$.
We assess the
importance of spiral arms by performing calculations with and without 
an additional 4 armed spiral component\footnote{Although we do not present the results here,
we also ran a calculation with a 2 armed spiral (with
$\epsilon=0.05$), and found the properties of the disc
(i.e. fractions of gas in different phases of the ISM, velocity
dispersion, star formation rate) very similar to those for run L5.}, using the potential from
\citet{Cox2002} and used in previous calculations,
e.g. \citealt{DBP2006}. As well as the galactic potential, self
gravity of the gas is also included. All the calculations also contain 
ISM heating and cooling, and stellar feedback. However we do not
include magnetic fields. All calculations use one million particles.

To set up our calculations, we assign gas particles a random
distribution, with velocities according to the rotation
curve of the galactic potential. We also assign the particles an
additional velocity dispersion by sampling from a Gaussian of 
width 6 km s$^{-1}$. In all calculations the gas lies within a radius
of 10 kpc, and the initial scale height is 200 pc. The total gas
  mass in the simulations is $2.5 \times 10^9$ M$_{\odot}$ in most cases, and $5
  \times 10^9$ M$_{\odot}$ for two calculations.
We include cooling and heating of the ISM, following the
prescription of \citet{DGCK2008}. Apart from feedback from
star formation, heating is mainly due to
background FUV, whilst cooling is due to a variety of processes
including collisional cooling, gas-grain energy transfer and recombination on grain surfaces. 
The particles are initially assigned a temperature of 700 K, but after
a few $10^7$ Myr, the gas develops a multiphase nature from 20 K to
$~2\times10^6$ K. 

Our prescription of stellar feedback involves inserting energy at
locations of star formation. The prescription does not involve any
timescale for the duration of star formation, or an efficiency per
free fall time, so it is independent of the Schmidt-Kennicutt
relation. However we do require a star formation efficency parameter, 
$\epsilon$, to provide a measure of the strength of the feedback, 
as we describe below. We test a range of values of $\epsilon$ in our calculations.

We include stellar feedback only when a pocket of gas becomes
sufficiently 
self-gravitating. This requires that i)
the density of a particle is greater than 1000 cm$^{-3}$, ii) the gas flow is converging, ii)
the gas is gravitationally bound (within a size of about 20 pc, or 3
smoothing lengths), iv) the sum of the ratio of thermal and rotation
energies to the gravitational energy is less than 1, and v) the total
energy of the particles is negative (see \citealt{Bate1995}). We test
whether gas satisfies these criteria after each (maximum particle) time
step, and if these
conditions are met, we assume that star formation occurs 
together with instantaneous supernova feedback. However unlike
\citet{Bate1995}, we do not include sink particles to represent stars
(or star clusters), rather we simply add an amount of energy to the
ISM, determined from the number of stars expected to form. 
The addition of this energy is generally
sufficient to disperse the gas. Thus from the gas
satisfying the above criteria, we determine the mass of molecular
gas. The time dependent molecular gas fraction for all the particles is calculated
according to \citet{DGCK2008}, using a rate equation which
includes the rate of H$_2$ formation on grains and a simplified
estimate of the photodissociation rate based on the local gas density.
The total energy from stellar feedback is then
\begin{equation}
 E_{SN}=\frac{\epsilon M_{H_{2}}} {160 M_{\odot}} 10^{51} \; ergs,
\end{equation}
where $\epsilon$ is the star formation efficiency. We assume that
each supernova contributes $10^{51}$ ergs of energy, and that one
supernova occurs per 160 M$_{\odot}$ of stars formed. The latter
assumes a Salpeter initial mass function with limits of 0.1 and 100
M$_{\odot}$.
Note however that although we refer to the feedback as `supernovae
feedback', we are simply adding energy to our calculations, which
collectively accounts for numerous processes including stellar winds,
radiation and supernovae.\footnote{\citet{Hopkins2011} perform similar
  style SPH calculations injecting energy into surrounding
  particles. However they use a more typical Kennicutt-Schimdt recipe,
  showing that the star formation rates they achieve are relatively
  independent of efficiency. Moreover they do not investiage cloud
  properties, which are more the focus of the present paper.}   
We add the energy according to a snowplough solution (see Appendix),
and deposit energy input instantaneously in the gas (approximately
$1/3$ thermal and $2/3$ kinetic energy).  

A summary of the different
calculations performed is provided in Table~1. We perform calculations with 2
different surface densities and star formation efficiencies between 1
and 40 per cent. The average surface density of the Milky Way is $\sim10$
M$_{\odot}$ pc$^{-2}$ \citep{Wolfire2003}.In order to include both
the cold phase of the ISM, and adequately resolve the Jeans length, requires
performing calculations with a number of particles which are currently
unfeasible.
Thus in the Appendix we also
show tests using different surface density thresholds, and adopting a
temperature floor of 500 K.

\begin{table}
\centering
\begin{tabular}{c|c|c|c|c|c|c|c}
 \hline 
Run & Surface density  & $\epsilon$ \\
& (M$_{\odot}$ pc$^{-2}$) & per cent \\
 \hline
L1 & 8 & 1 \\
L5 & 8 & 5 \\
L10 & 8 & 10 \\
L20 & 8 & 20 \\
L40 & 8 & 40 \\
M5 & 16 & 5 \\
M10 & 16 & 10 \\
L5$_{nosp}$ & 8 & 5 \\
L10$_{nosp}$ & 8 & 10 \\
\hline
\end{tabular}
\caption{In the above table we list the calculations presented in this
  paper. In Runs L5$_{nosp}$
  and L10$_{nosp}$ the spiral component of the potential is not
  included. $\epsilon$ is the star formation efficiency (see text).}
\label{runs}
\end{table}

\section{Global evolution of the disc}
In most of our calculations, where there is sufficient energy feedback, 
the evolution of the disc follows a
similar pattern. So we provide here an overall picture of the typical 
evolution. The gas in the disc initially cools, and in
the case of the spiral potential, gas is gathered into clouds by the
spiral shocks. Once the gas is sufficiently cool and dense, widespread star
formation occurs in the disc. This star formation generates a
substantial amount of warm gas, which then prevents star formation
occurring at the same rate. The gas then slowly cools, and the star
formation rate slowly increases again. This behaviour is somewhat episodic, but
after the first burst of star formation, the distributions of cold and
warm gas, 
and the star formation rate, tend
to be much less extreme. In some cases, we reach an equilibrium state,
but in some calculations this is not feasible, as we shall explain. The distribution of the ISM, and star
formation rates naturally depend on the parameters of our calculations,
which we discuss in detail in the following sections.

\section{Results - The star formation efficiency}
First we show results where we compare different star formation efficiencies.
The star formation efficiency ($\epsilon$) controls the amount of stellar feedback
(i.e. energy) that is included in our simulations. We used five
different values of $\epsilon$, 1, 5, 10, 20 and 40 per cent. Observed
estimates are from a few to 10 or 20 per cent \citep{Evans2009}. Our
value of 40 per cent may be unrealistically high except perhaps for
starbursts, whilst for the 1 per
cent case, we find feedback has very little effect.  
\begin{figure}
\centerline{
\includegraphics[scale=0.52, bb=0 20 450 1200]{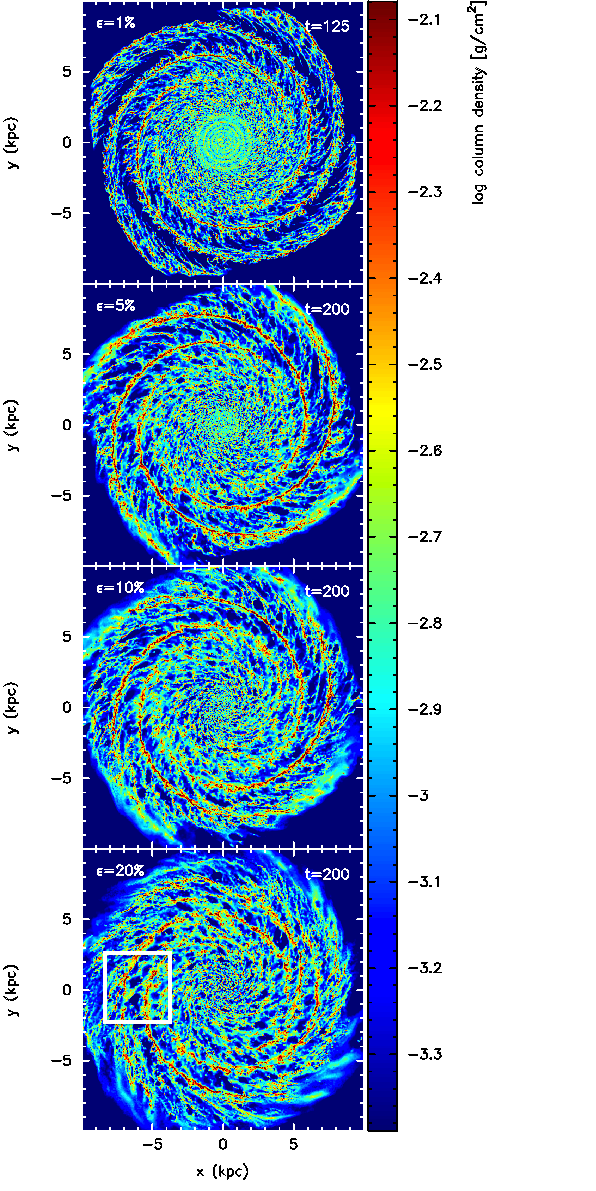}}
\caption{The gas column density is shown for calculations with star
  formation efficiencies of 1, 5, 10 and 20 per cent (Runs L1, L5, L10 and
  L20) at a time of 125 Myr for the 1 per cent case (top),
  and 200 Myr in the other panels.}
\end{figure}

\begin{figure}
\centerline{
\includegraphics[scale=0.28]{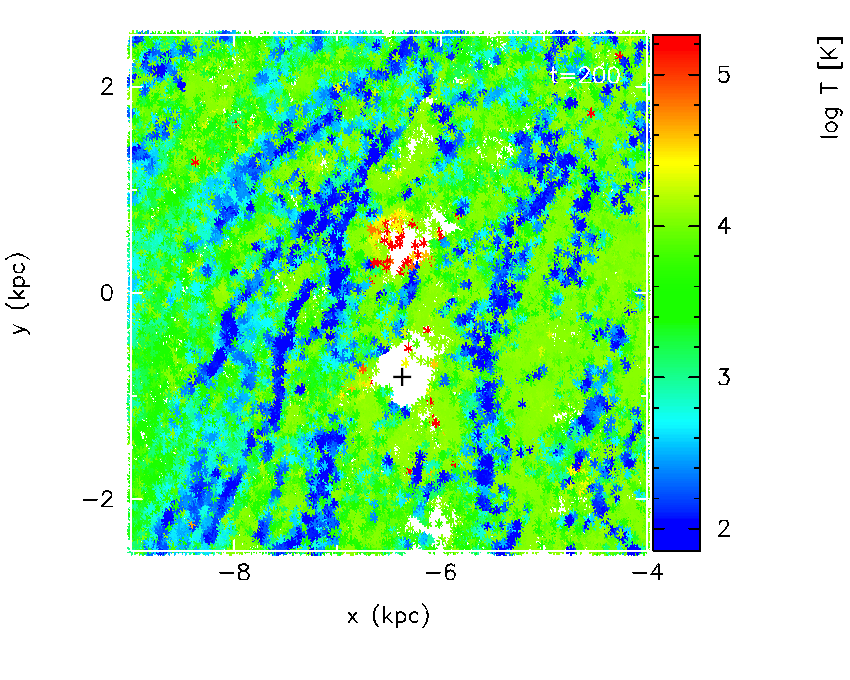}}
\caption{Particles are plotted for a subsection of Run L20 (with 20 per
  cent efficiency), specifically illustrating holes that have been
  blown out by supernovae. The region selected is indicated by the
  white box in Fig.~1. All the particles are plotted which lay
  between
  $z=-0.5$ and 0.5 kpc, their colour indicates the
  temperature.}
\end{figure}

\subsection{Structure of the disc}
We show the structure of the disc (including the spiral component of the
potential) for efficiencies of 5, 10 and 20 per cent at a time of 200
Myr, and 1 per cent at a time of 125 Myr in Fig.~1. For the 1, 5 and 10 per cent
cases, the 4-armed spiral structure is clear, but there is also a lot of
substructure in the disc, and many filamentary interarm features, or
spurs.

For the 1 per cent case, much of the gas becomes confined to dense,
gravitationally bound clumps either in the spiral arms, or moving from
the spiral arms into the interarm regions (see also
\citealt{Dobbs2011}). As such, we are only able to run this
calculation for a short time. Because
the gas is gravitationally bound in these dense clouds, the elongated
spurs seen in the other panels are unable to form as readily. The
structure for the 5 and 10 per cent cases is similar to what we would
expect to see in spiral galaxies -- there are clear spiral arms and
spurs. Thus the feedback does not prevent the formation of dense
clouds of gas, nor the evolution of these clouds into spurs.
For the 20 per cent efficiency case, feedback begins to
dominate the structure of the disc. The arms are more broken
up, and consequently it is more difficult to distinguish spiral arms
from interarm substructure. The structure for the 40 per cent case is
similar to that for 20 per cent, though the spiral arms are even less clear.
The interarm spurs are clearest for the 5 per cent case.

In the 20 per cent efficiency case, there are many
large holes in the disc, which have been blown out by supernovae.
Such holes occur for all the $\epsilon>1$ per cent cases, but are less
noticeable with lower efficiencies. It is also not so obvious to determine
which features are holes blown out by supernovae, and which features
are simply less dense regions between spurs or clouds which occur regardless of
supernovae feedback (e.g. \citealt{Dib2005,Dobbs2006,Dobbs2008}). In Fig.~2 we
show a subsection of the 20 per cent efficiency case (Run L20),
corresponding to the box in Fig.~1. We plot the temperature of all
particles for $|z|<0.5$ kpc in the region. The two holes in the
centre, both several 100 pc across,
are clearly associated with hot gas, and therefore supernovae
feedback. The lower hole (as indicated by the cross) is particularly
devoid of any particles.

\begin{figure}
\centerline{
\includegraphics[scale=0.56, bb=30 280 350 880]{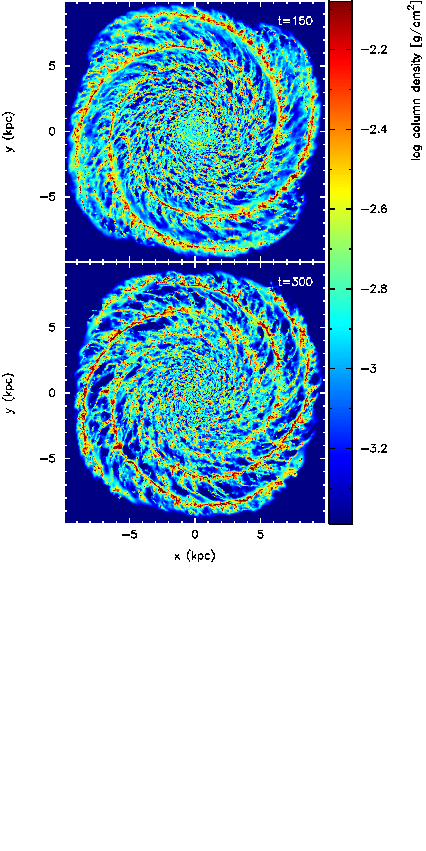}}
\caption{The gas column density is shown for the calculation using a star
  formation efficiency of 5 per cent (Run L5), at times of 150 and 300 Myr.}
\end{figure}
We show the structure of the disc for the 5 per cent efficiency calculation
(Run L5) at times of 150 and 300 Myr in Fig.~3. At 150 Myr, the arms are slightly
wider, and there is a smaller arm-interarm density contrast.
At 300 Myr, there are a few more noticeable, and denser, clumps along
the arms, but the structure is largely similar. 
For the highest star formation efficiency simulations
the structure is more time dependent.
As we will show in Section~4.5, the star formation rate peaks
between 50 and 100 Myr. For the
higher efficiency cases (Runs L20 and L40) this leads to substantial
disruption of the gas disc (particularly for Run L40, where holes of 
a few kpcs across are produced) at earlier times.

\subsection{The CNM, WNM and unstable regime}
\begin{figure}
\centerline{
\includegraphics[scale=0.4]{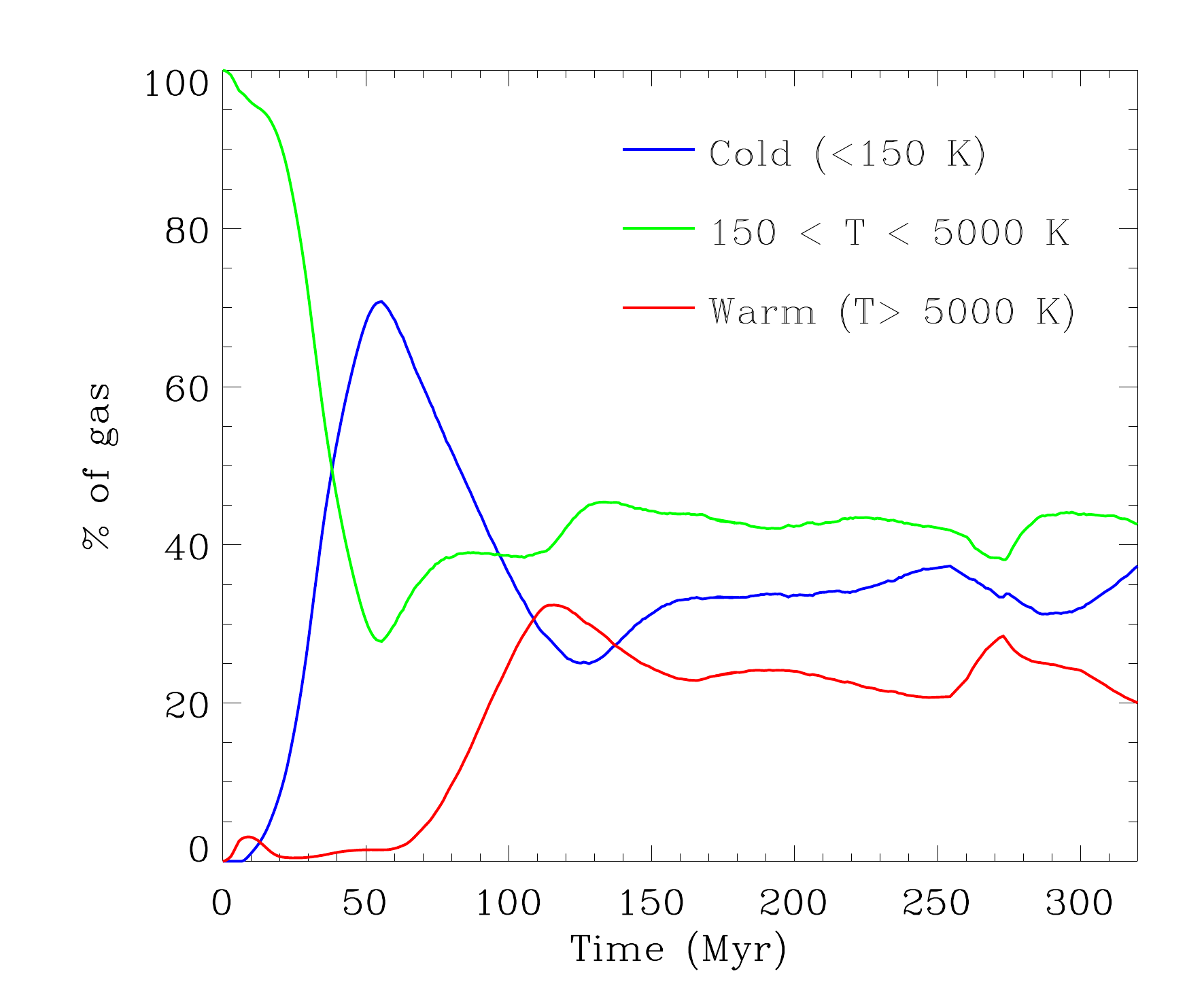}}
\centerline{
\includegraphics[scale=0.4]{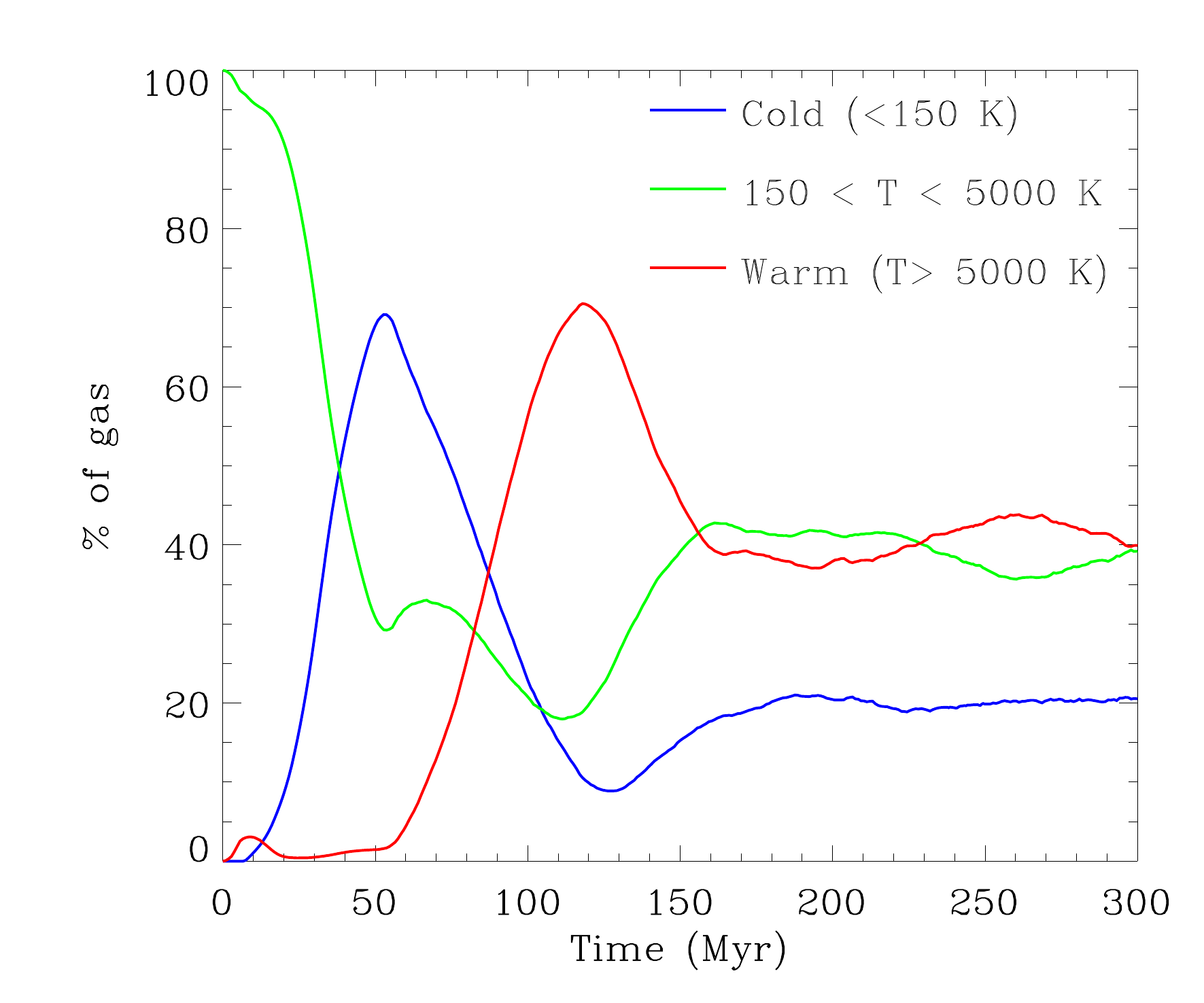}}
\caption{The fraction of gas in the cold, unstable and warm phases is
  shown versus time for Runs L5 and L20, the calculations with $\epsilon=$ 5 per cent (top), and
  20 per cent (lower).}
\end{figure}

\begin{figure}
\centerline{
\includegraphics[scale=0.4]{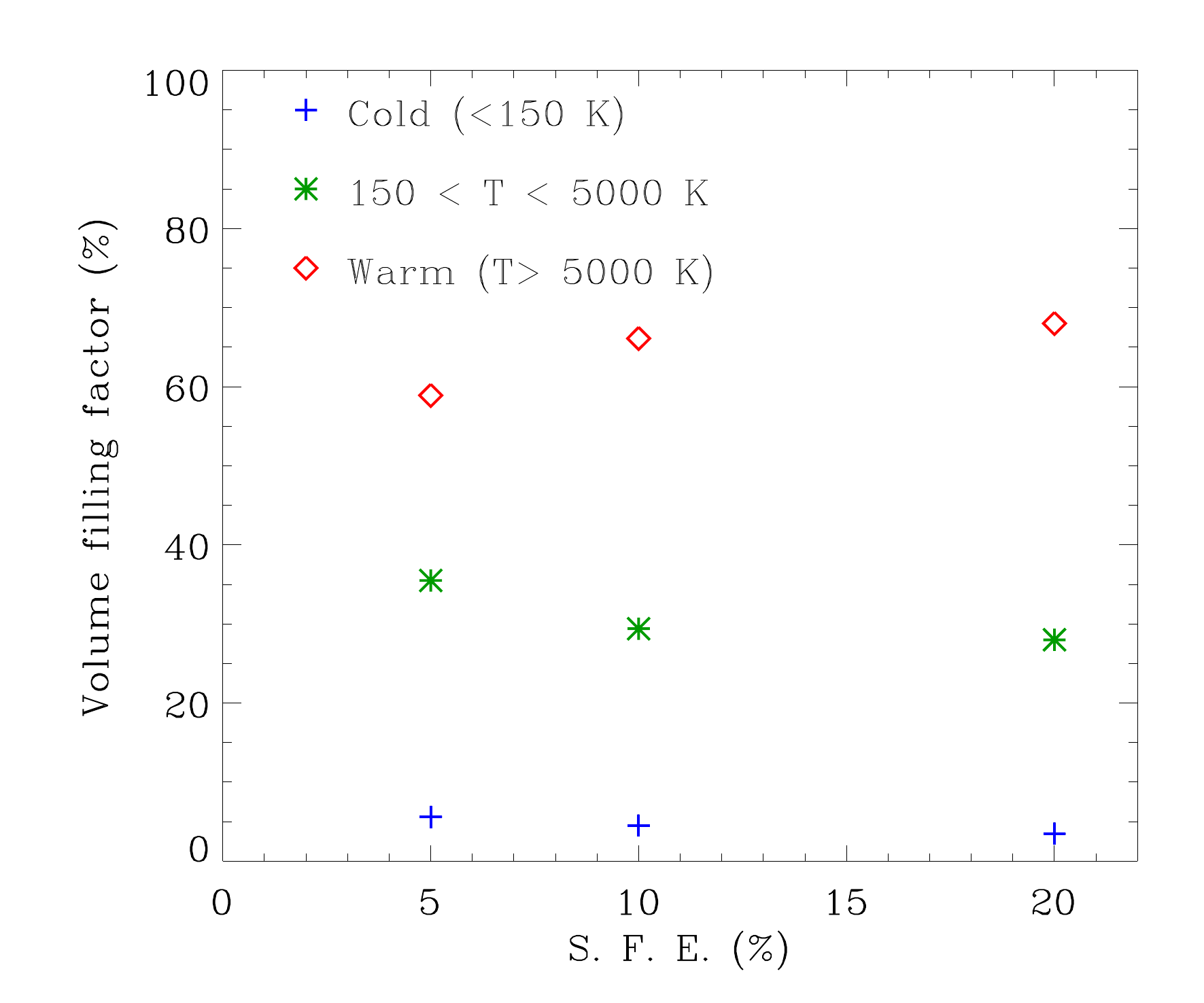}}
\caption{The volume filling factor for the cold, unstable and warm phases is
  plotted for the calculations with efficiencies of 5, 10 and 20 per
  cent (Runs L5, L10 and L20), at a time of 200 Myr.}
\end{figure}
The structure of the cold (i.e. $\sim 100$ K) and warm ($\sim
10^4$ K) phases of the ISM as shown in Fig.~2 is typical for these
calculations. The cold gas is largely situated along the spiral arms
and spurs, with the warmer gas filling in the regions between, thus
similar to \citet{DGCK2008}. In the lower efficiency cases there is a
little more cold gas, and the spiral arms are more continuous. The
hotter gas (not included in \citealt{DGCK2008}) is largely confined to
supernovae bubbles and gas outside the plane of the disc.

In Fig.~4, we show the fraction (by mass) of gas in the cold, unstable and warm
phases of the ISM.
For the calculation with $\epsilon=$ 5 per cent, approximately one
third of the gas lies in each phase. This is not dissimilar to
observations and other simulations
(e.g. \citealt{Heiles2003,Gazol2001,KKO2008}). 
When $\epsilon=$ 20 per cent, only 20 per cent of the gas is in the cold phase,
which is probably too low to reconcile with observations. We only obtained a
high fraction 
of cold gas (65 per cent), similar to calculations without feedback
\citep{Dobbs2008} in the calculation with 1 per cent efficiency.
There is some gas in the hot phase in these calculations but this
constitutes very little mass, $\lesssim 1$ per cent of the total gas in the
calculations with 5 and 10 per cent efficiency and a few per cent in
the calculations with 20 and 40 per cent efficiency. For the 40 per
cent efficiency case (Run L40), the majority of the gas (60--70 per
cent) lies in the warm phase.

The volume filling factor for the different phases is shown in Fig.~5,
for the calculations with 5, 10 and 20 per cent efficiency, at a time
of 200 Myr. We
calculate the filling factor for gas within a height of 1 kpc, and a
radius of 10 kpc. There is
little variation with efficiency, the warm and cold gas occupying
slightly higher and lower volumes for higher star formation efficiencies. 
Volume filling factors are not well known for the ISM, though
the filling factor of the CNM is believed to be 1 \citep{Cox2005} or a
few \citep{McKee1977} per cent. The filling factor of the WNM is
thought to be about 50 per cent \citep{Heiles2003}. 
Thus our estimates are in approximate agreement with the observations.

\subsection{Velocity dispersion of the gas}
\begin{figure*}
\centerline{
\includegraphics[scale=0.4]{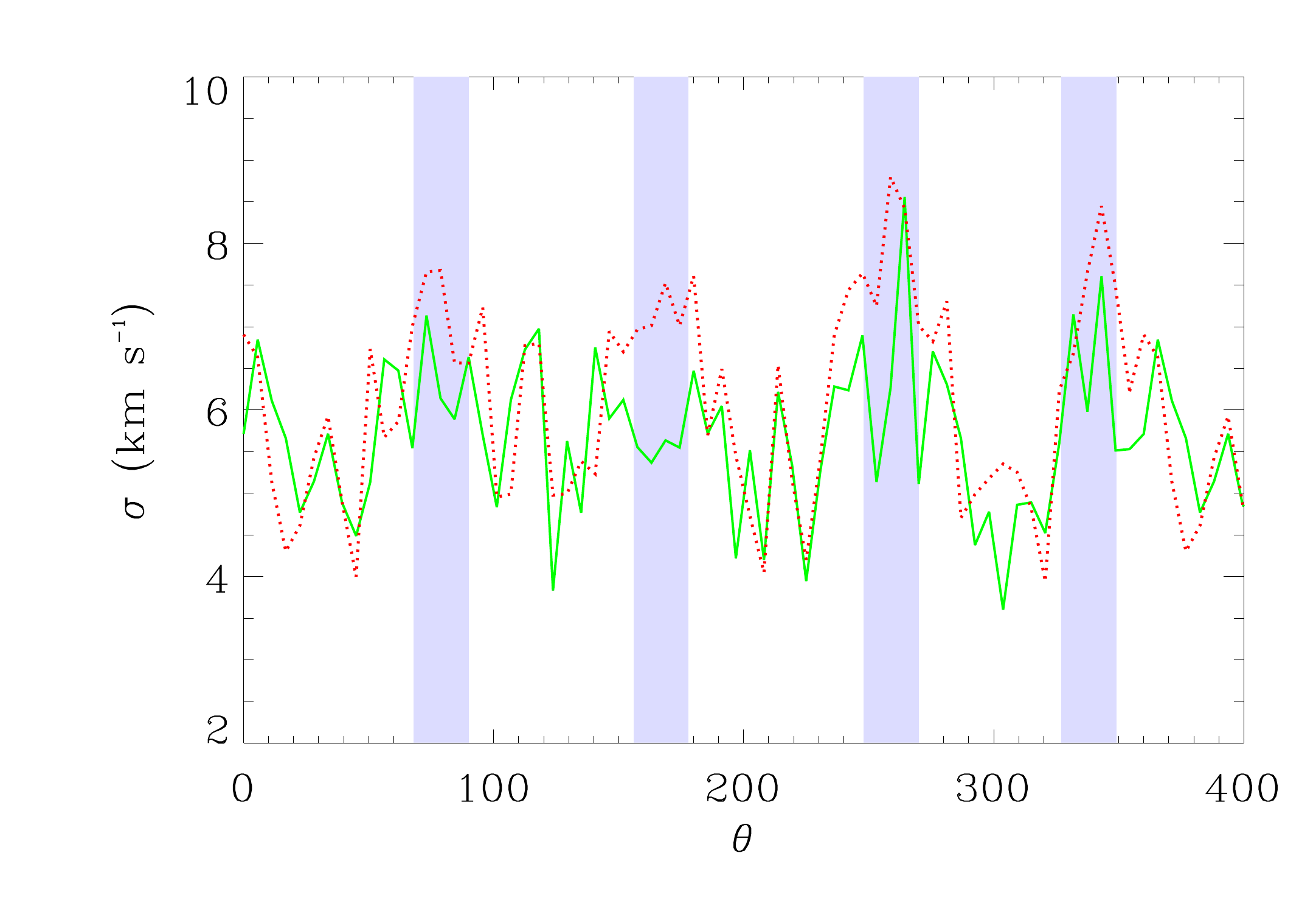}
\includegraphics[scale=0.4]{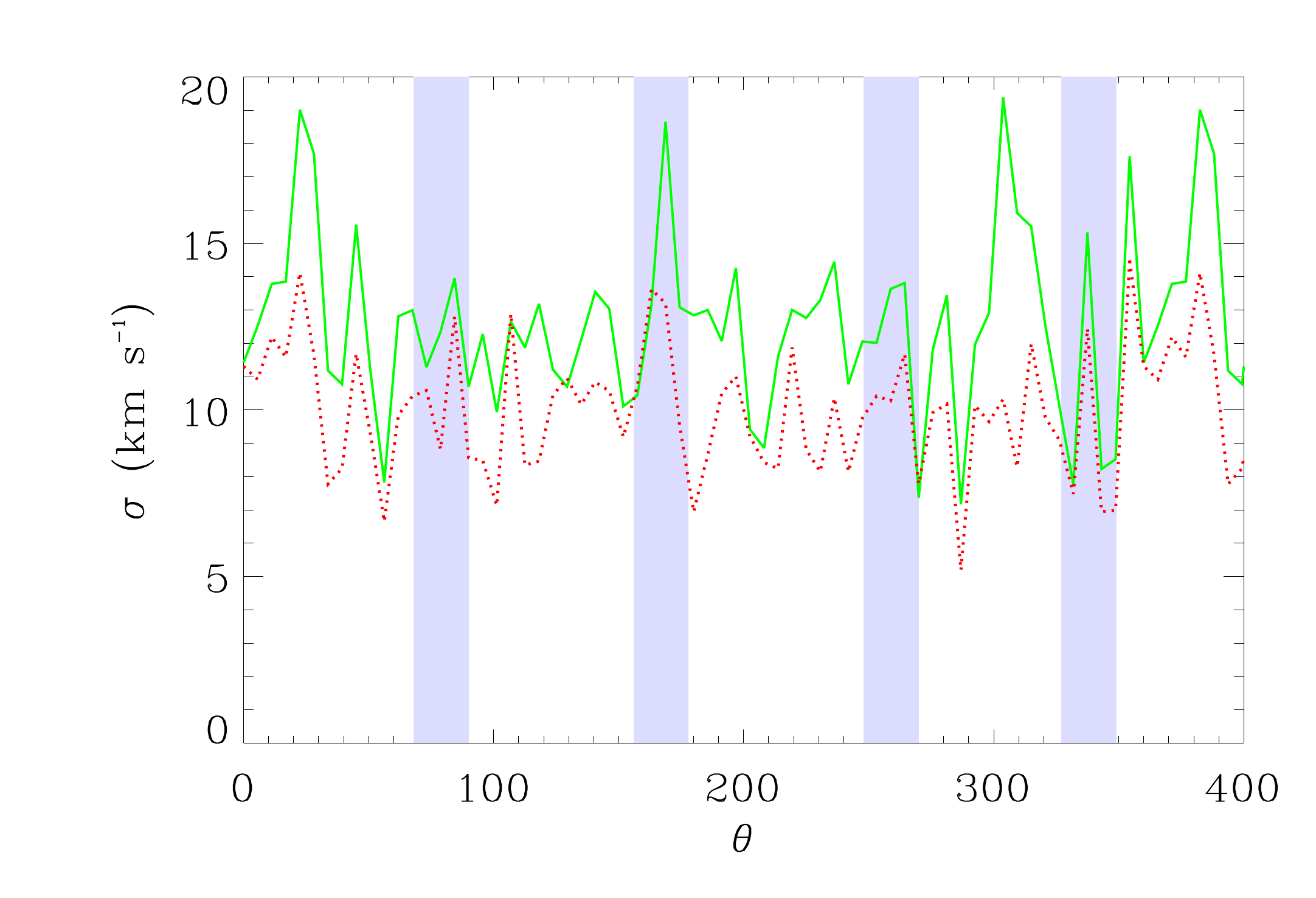}}
\caption{The velocity dispersion is calculated for an annulus of the
  disc for the 5 (left) and 20 (right) per cent efficiency
  calculations (Runs L5 and L20) at a time of 200 Myr. $\theta$ is the direction anticlockwise round the
  disc. The grey regions indicate the location of the spiral arms. The
velocity dispersions, $\sigma_r$ (red, dotted) and $\sigma_z$ (green,
solid) are between 4 and 8 km s$^{-1}$ when $\epsilon=$5 per cent,
and between 6 and 20 km s$^{-1}$ when $\epsilon=$20 per cent.}
\end{figure*}
Observations indicate that, with the exception of very hot gas, the
velocity dispersion in the ISM is typically 5--10~km~s$^{-1}$. 
In previous work without stellar feedback \citep{DBP2006}, we found that a
velocity dispersion of this magnitude is achieved in the spiral arms,
but quickly decays in the interarm regions. In Fig.~6, we show the
velocity dispersion from an annulus of the disc, from the calculations
with 5 (left) and 20 (right) per cent efficiencies (Runs L5 and L20), at a time of 200 Myr. We select
particles within an annulus of width 1 kpc, a central radius of 7.5
kpc, and divide the annulus into 64 sections azimuthally. We show the
velocity dispersion in the plane of the disc
($\sigma_r=\sigma(\sqrt{v_x^2+v_y^2}))$\footnote{Note that as we 
  select gas away from the centre, where the rotation curve is flat,
  there is little difference in the velocity dispersion if
  we subtract the rotational velocities first.} and the
vertical dispersion, $\sigma_z$.

We see from Fig.~6 that for the 5 per cent efficiency case (L5),
there tend to be
peaks in the velocity dispersion corresponding to the location of the
spiral arms. The peaks are also higher for the velocity in the plane
of the disc, whilst in the interarm regions, the velocity dispersions
are similar in the two directions. However unlike \citet{DBP2006}, the
velocity dispersion is still maintained above 4 km s$^{-1}$ in the
interarm regions. The enhancement of the velocity
dispersion in the spiral arms is likely due to the higher star
formation rate in the spiral arms, but the relatively higher increase for the
dispersion in the plane of the disc suggests the spiral shock is also
relevant. In contrast, the velocity dispersion for the 20 per cent
efficiency case (Run L20), there is no correlation of the dispersion
with the spiral arms. This is perhaps not surprising given that the
spiral structure is weaker in this case. The velocity dispersion is
observed to be higher in the spiral arms of NGC 0628
\citep{Shostak1984} and M51 (\citealt{Hitsch2009}, Fig.~11), but not
NGC 1058 \citep{Petric2007}, though the latter has a less prominent
spiral structure. The mean ratio $\sigma_z/\sigma_r$ is 0.9 for the 5
per cent case, and 1.2 for the 20 per cent calculation. These values
decrease if only gas close to the midplane is selected.

We find $\sigma_r$ and $\sigma_v$ take values
of 4--8 km s$^{-1}$ when $\epsilon=5$ per cent ($\bar{\sigma_z}=5.7$ km
  s$^{-1}$), 5--16 km s$^{-1}$ when $\epsilon=10$ per cent ($\bar{\sigma_z}=9.1$ km
  s$^{-1}$) and 8--20  km s$^{-1}$ when $\epsilon=20$ per cent ($\bar{\sigma_z}=12.6$ km
  s$^{-1}$). Thus the velocity dispersions when $\epsilon = 5-10$ per
  cent roughly match the observations, but are slightly high for
  $\epsilon =20$ per cent.

We also calculated the velocity dispersions in the cold, unstable and
warm phases. For the $\epsilon=5$ per cent calculation,
$\bar{\sigma_z}$ is 5.1, 6 and 7.3 km s$^{-1}$ for the cold, unstable
and warm phases respectively. When $\epsilon=20$ per cent,
$\bar{\sigma_z}$ is around 12 km s$^{-1}$ in the cold and unstable
phases, and 20 km s$^{-1}$ in the warm phase. For Run L5,  
the ratio $\sigma_z/\sigma_r$ is lower for the cold gas, and there is a greater
tendency for the $\sigma_z$ and $\sigma_r$ to coincide with the
spiral arms for the cold gas. The latter behaviour is also seen in
M51, when comparing HI and H$_2$, and in fact the peak HI dispersion is not at
all coincident with the spiral arms \citep{Hitsch2009}.

In all cases, the velocity dispersions vary with time earlier in the
calculations, following the variation in the star formation rate
(see Section~4.5). The velocity dispersions are higher when the  star
formation rate is higher.

\subsection{Scale heights of the gas}
\begin{figure}
\centerline{
\includegraphics[scale=0.35]{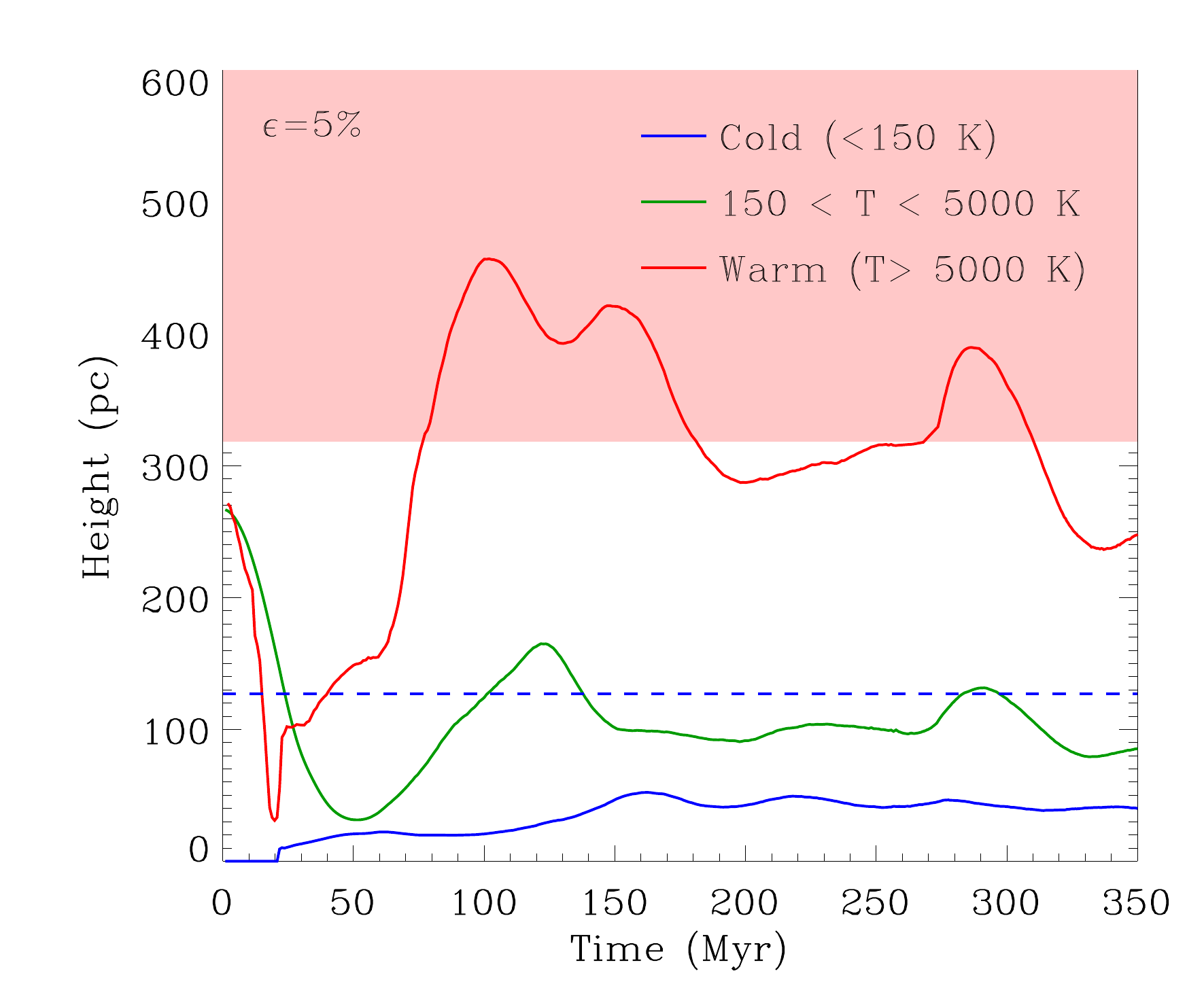}}
\centerline{
\includegraphics[scale=0.35]{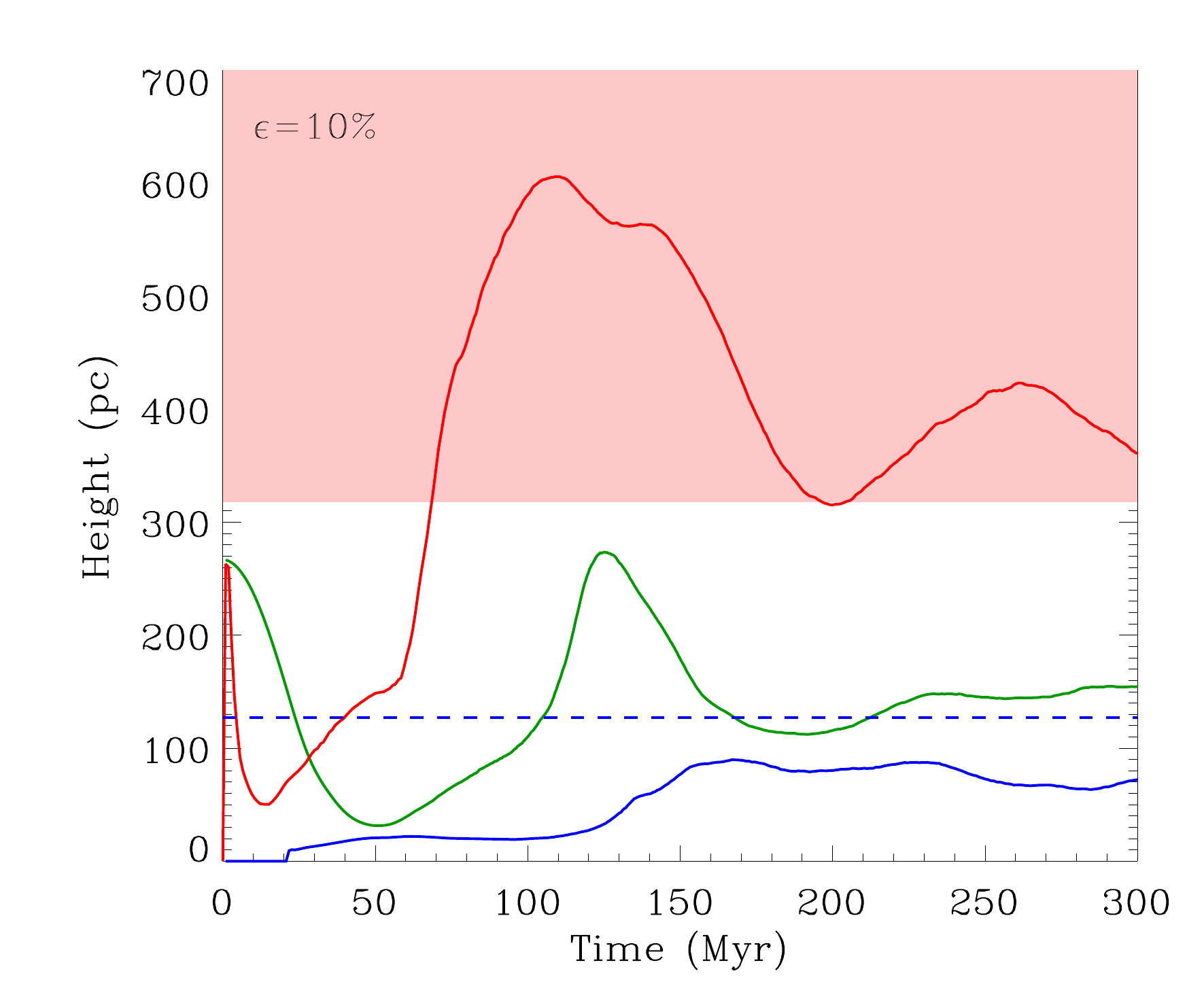}}
\centerline{
\includegraphics[scale=0.35]{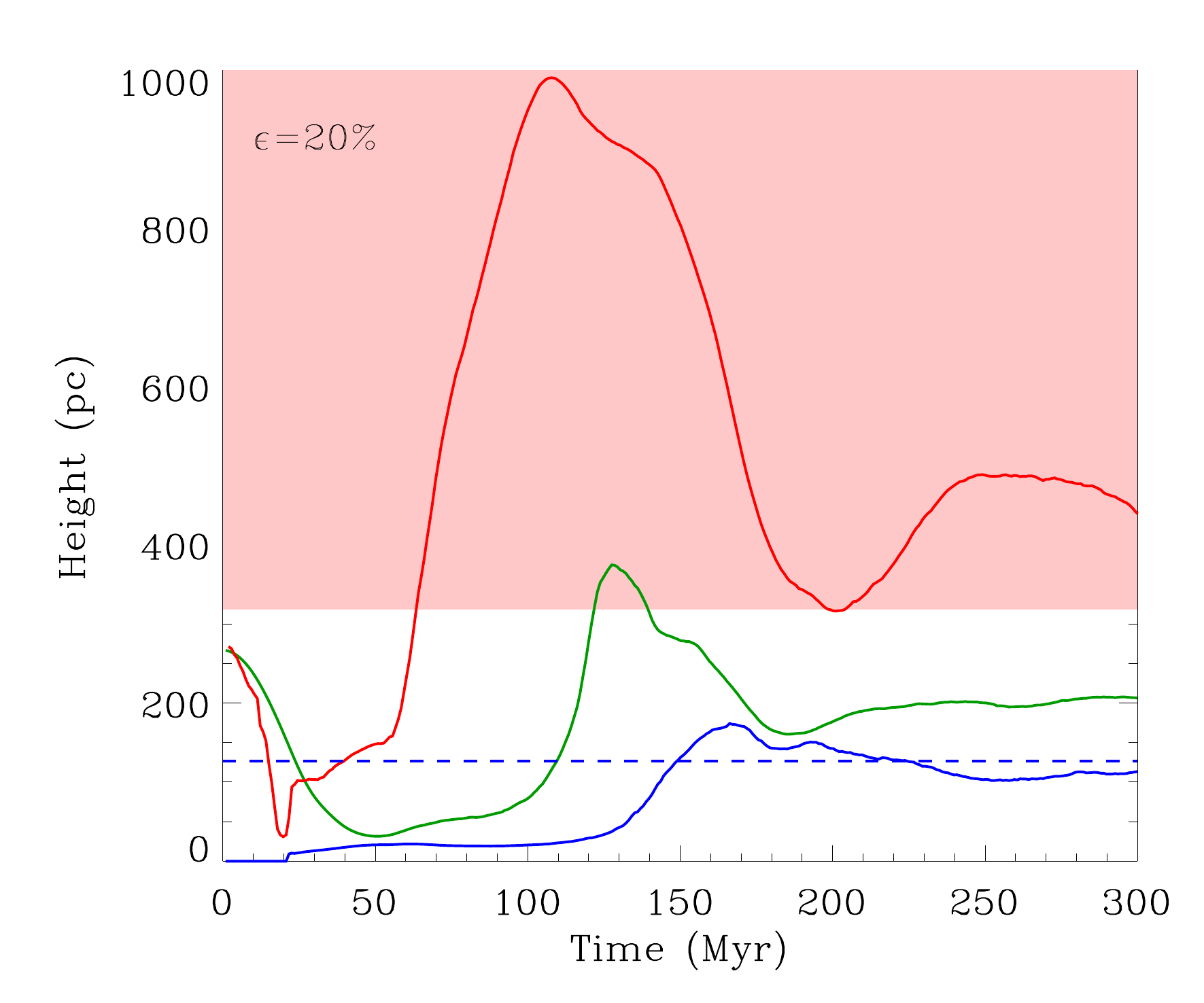}}
\caption{The scale heights of the cold, unstable and warm phases of
  the ISM are shown with time for the 5 (top), 10 (centre) and 20
  (lower) per cent efficiency calculations (Runs L5, L10 and L20). The
  scale height of the warm gas can be between 325 pc and 1 kpc (red
  filled region), depending on whether the gas is ionised, whilst the
  scale height of the cold gas is estimated to be 127 pc (blue dashed line) \citep{Cox2005}.}
\end{figure}

\begin{figure}
\centerline{
\includegraphics[scale=0.43, , bb=-100 20 330 330]{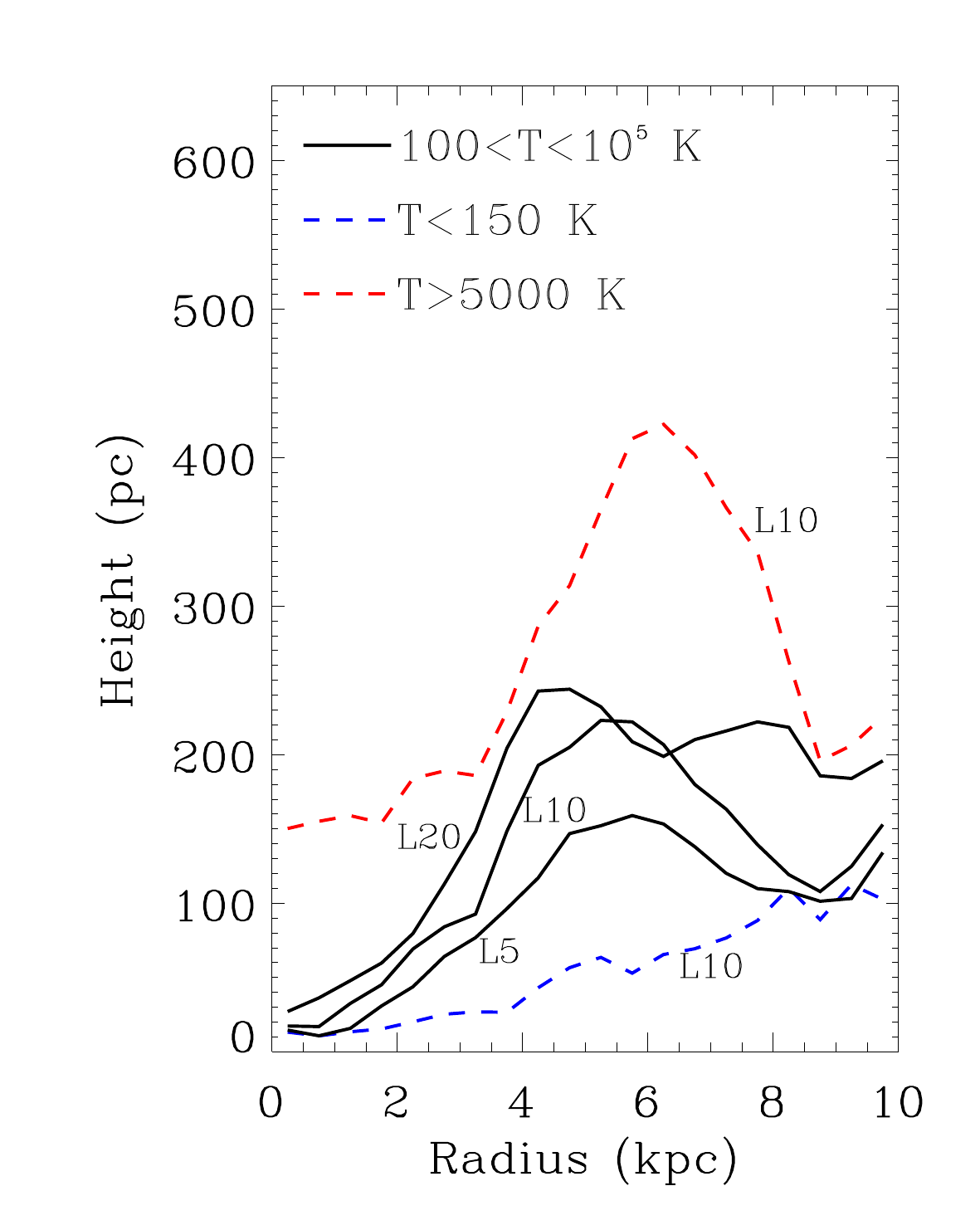}
\includegraphics[scale=0.56, , bb=0 0 330 300]{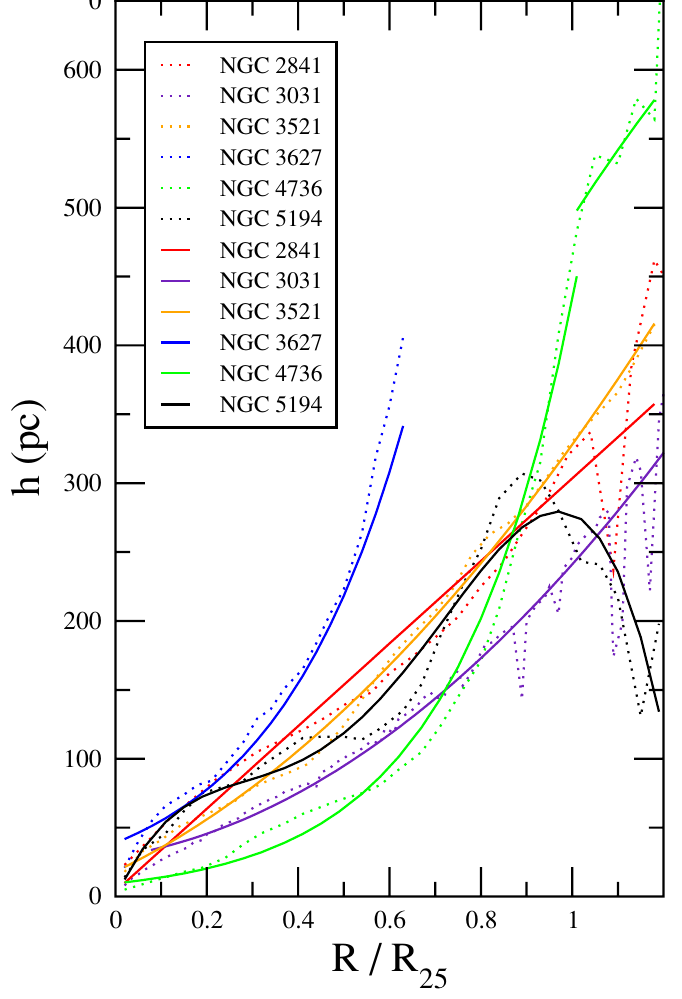}}
\caption{The scale height versus radius is shown for the
  calculations L5, L10 and L20 as indicated. The scale height is
  calculated from gas between 100 and $10^5$ K. 
Scale heights for the cold gas, and warm gas, from the
  10 per cent efficiency calculation (Run L10) are shown separately as
  blue and red dashed lines. The right hand panel shows the scaleheight of neutral HI for a
selection of early-type spiral galaxies \citep{Bag2011}. The solid and
dotted lines represent different techniques. In both
simulations and observations
the scale height is much lower in the inner regions of the disc.}
\end{figure}

\begin{figure}
\centerline{
\includegraphics[scale=0.27, bb=200 -50 650 750]{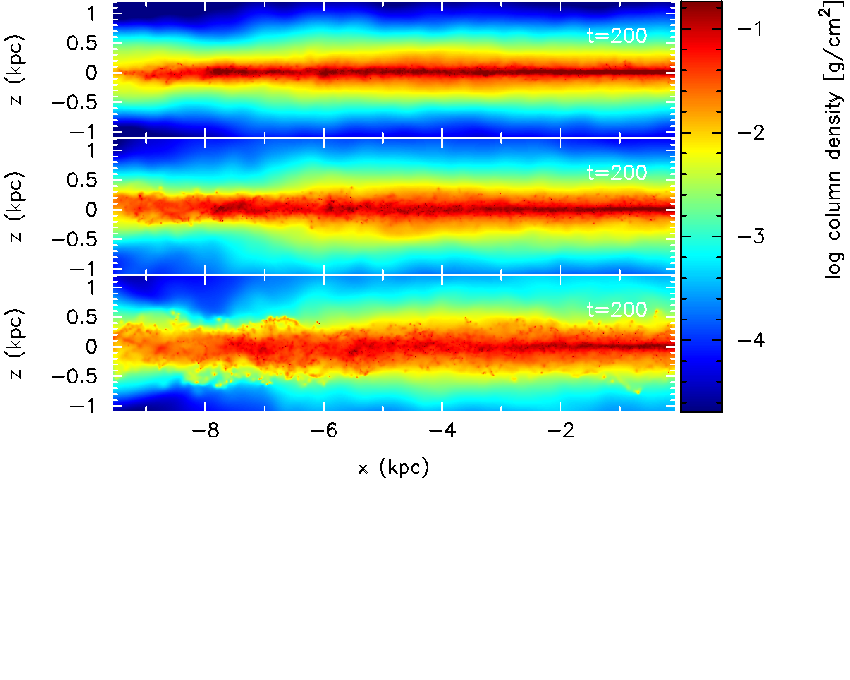}}
\centerline{
\includegraphics[scale=0.27, bb=200 0 650 150]{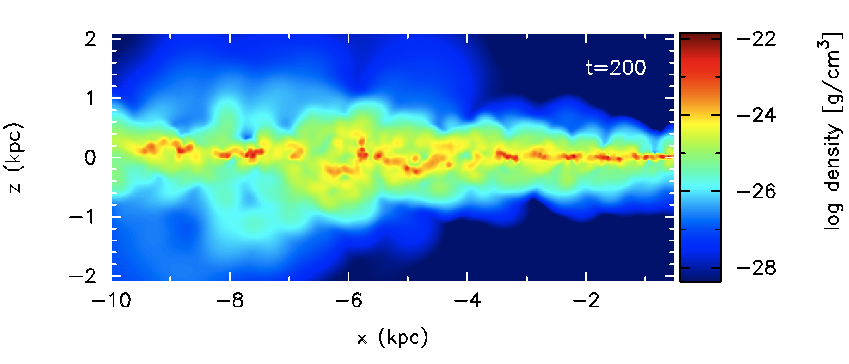}}
\caption{The column density in the vertical plane is shown (top panel)
  of the 5 (top-most), 10 (center) and 20 (third panel) per cent
  efficiency at time of 200 Myr. The column density is calculated
  by integrating along the direction of the $y$-axis. 
Without any feedback, the vertical extent is about half
  that of the 5 per cent (top) case, thus feedback is essential to
  reproduce the disc scale height. The lowest panel shows a cross
  section in the $zx$ plane for the 10 per cent efficiency case.}
\end{figure}
The fact that we find velocity dispersions of order 5--15~km~s$^{-1}$
in these models suggests that we are reproducing the dynamics of the
ISM reasonably well. The scale heights of the disc are then largely
determined by the galactic potential we adopt.

In Fig.~7, we show the scale heights for the gas in the different
phases of the ISM with time for the 5, 10 and 20 per cent
calculations. As there is a substantial variation of scale height with
radius, as shown in Fig.~8, we calculate the average scale height for gas
only between
radii of 5 and 10 kpc. 
The scale heights of the gas components
suggested by \citet{Cox2005} are 127 pc for the cold gas, and the warm
gas can have heights between 325 pc and 1 kpc depending on whether the gas is
ionised (see Fig.~7). From Fig.~7, the 20 per
cent efficiency model appears to agree best with the observations for the cold gas. 
Though if we take a radius of 8 kpc, which is similar to the solar
neighbourhood, the 10 per cent efficiency case
best fits the observations, with the scale height of the cold gas
around 100 pc (Fig.~8).

We also show in Fig.~8 the observed scale
heights versus radius for neutral HI in nearby spiral galaxies, from the THINGS
survey \citep{Bag2011}. Both these and the simulations show an
increase in the scale height with radius, from $<50$ pc to $>200$ pc,
although our scale heights tend to flatten or decrease at the largest
radii. The variation with radius in our simulations is predominantly
due to the galactic potential we use.

We find an
offset between the time supernovae first occur (around 40 Myr) and the
increase in scale height for the different phases (Fig.~7). This offset is
largest for the cold gas, where it is around 100 Myr. We interpret
the offset for the warm gas as the time it takes for gas to leave the
plane of the disc, i.e $z_h/c_s$ where $z_h$ is the scale height and
$c_s$ is the sound speed. For the cold gas, we expect cumulative
effects from star formation over a longer period of time are also
required to increase the scale height.

We further illustrate the impact of supernovae on the vertical
structure of the disc in Fig.~9. Before stellar feedback occurs, 
the gas is confined to a very narrow ($\lesssim 40$ pc) band
in the mid-plane. For the 5 per cent efficiency case, gas is still
fairly confined to the mid-plane, but the scale height is around
double, and the material is less continuous in the mid plane. For the
20 per cent case, there is a lot more vertical structure in
the gas. The lowest panel shows a cross section through the disc for
the 10 per cent efficiency case after 200 Myr. Though we do not show the 5 and 20
per cent cases, for the 5 per cent case, the gas is
more continuous in the midplane (which is actually more similar to the
Canadian Galactic Plane (HI) Survey \citep{Douglas2010}), whilst the 20 per cent case, there is
more vertical structure. 
\begin{figure}
\centerline{
\includegraphics[scale=0.3, bb=20 0 500 350]{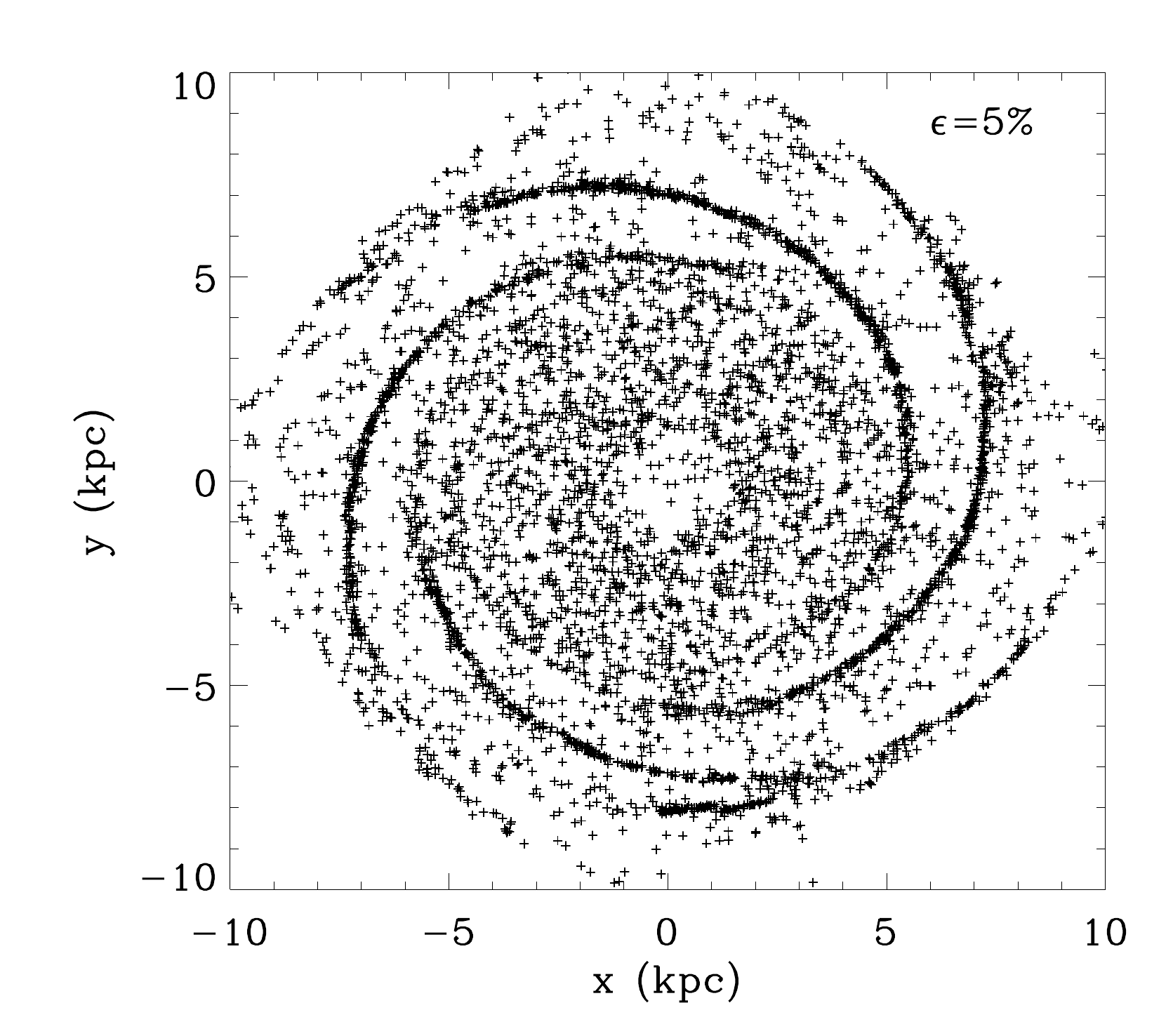}
\includegraphics[scale=0.3, bb=50 0 500 400]{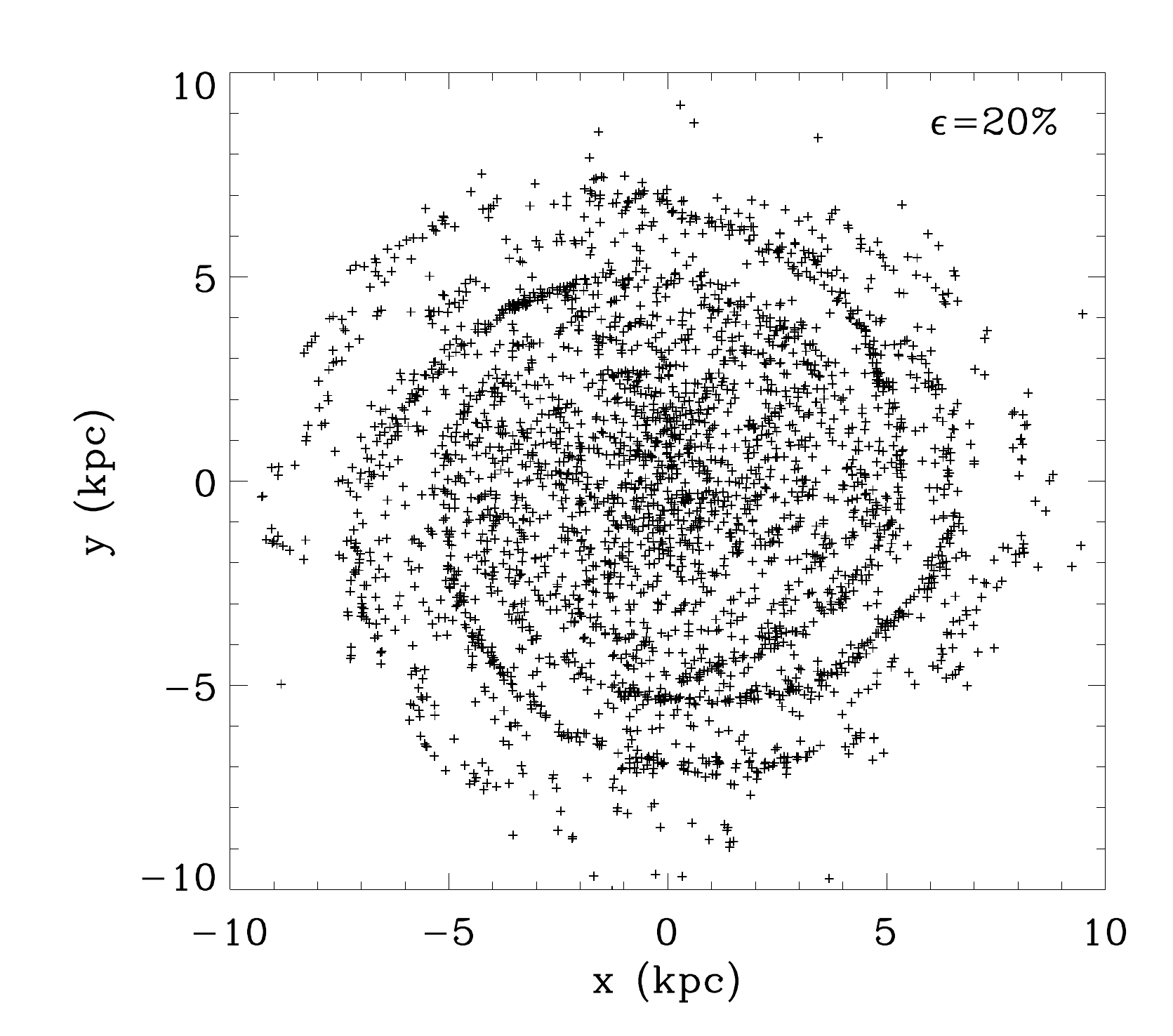}}
\centerline{
\includegraphics[scale=0.22]{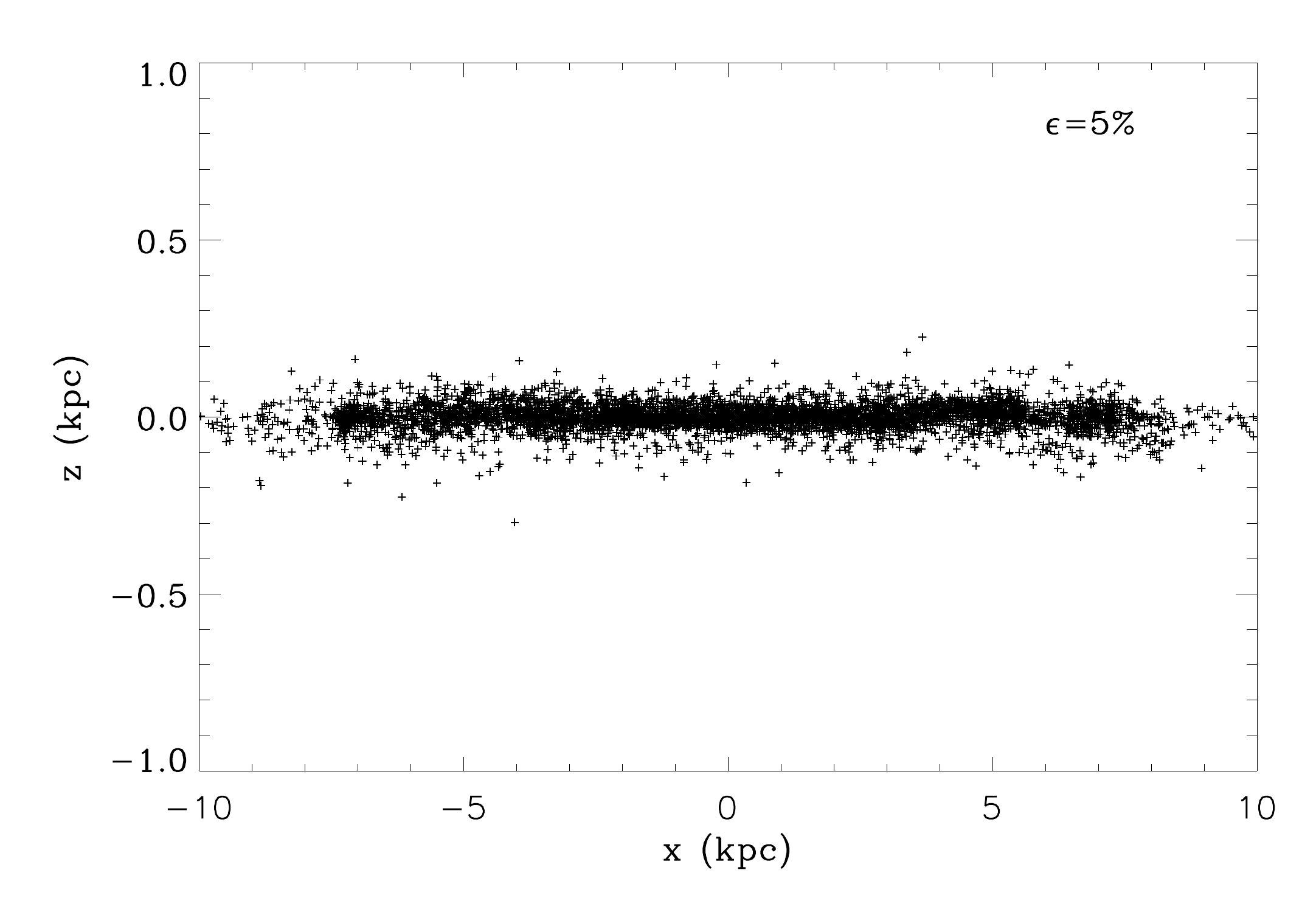}
\includegraphics[scale=0.22]{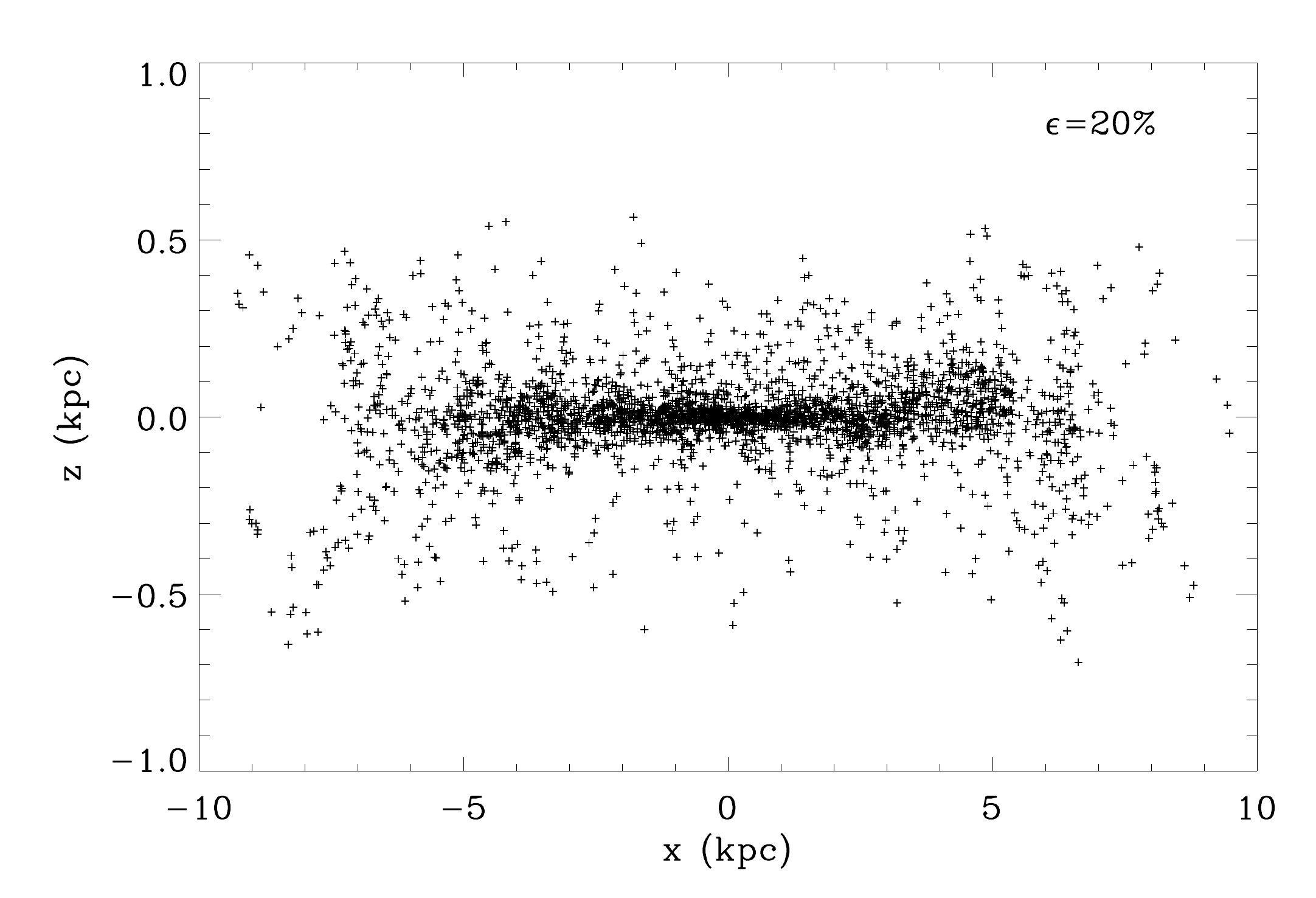}}
\caption{The distribution of star formation events is displayed over a
  20 Myr period (subtracting the phase of the potential) for the 5 (left) and 20 (right) per cent efficiency
  calculations (L5 and L20), in the plane of the disc (top), and out of the plane (lower).}
\end{figure}


As noted earlier, we would obtain different (and possibly
  higher) scale heights if we used a different
potential.  We also note that we do not include magnetic or cosmic ray pressure,
which would increase the scale height of the warm components, and
perhaps indirectly the cold gas. Neither do we allow for a warp which
is present in our, and other galaxies. 
In addition, the scale height may not be the most useful measure of
the vertical structure of the disc, for example if most of the gas is
confined to the central midplane, whilst other clouds lie at
far outside the midplane. 
\citet{Douglas2010} produced synthetic HI observations of gas from
previous simulations. Similar
synthetic HI maps, using the outcomes from the simulations presented
here with feedback, may indicate better which of our models best
reproduces the vertical structure of the HI in galaxies such as the
Milky Way. Figs.~7, 8 and 9 however suggest we would
obtain a much better agreement with the observations compared to the 
calculations without feedback shown in \citet{Douglas2010}, where the
gas was clearly too confined to the midplane.

\subsection{Distribution and scale heights of supernovae}
In Fig.~10 we display the locations of supernovae events over a 20 Myr
period (from 180-200 Myr) for the calculations with 5 and 20 per cent
efficiencies. The supernovae are concentrated to the spiral arms,
especially for the 5 per cent case, as expected
\citep{McMillan1996}. The supernovae occupy a much
larger vertical distribution for the 20 per cent case, compared to the
5. The scale heights of the supernovae events are 40 and 80 pc for the
5 and 20 per cent calculations respectively. This compares well 
to the distribution of OB stars, which are found to
have scale heights of around 
30--60 pc \citep{Mihalas1981,Reed2000,Maiz2001,Urquhart2011}.
Scale heights of Type II supernovae are likely to be larger than those
in our models \citep{Ferriere2001}, partly as the feedback is
instantaneous, although estimates based
on observations are also highly uncertain.

\subsection{Star formation rates}
\begin{figure}
\centerline{
\includegraphics[scale=0.42]{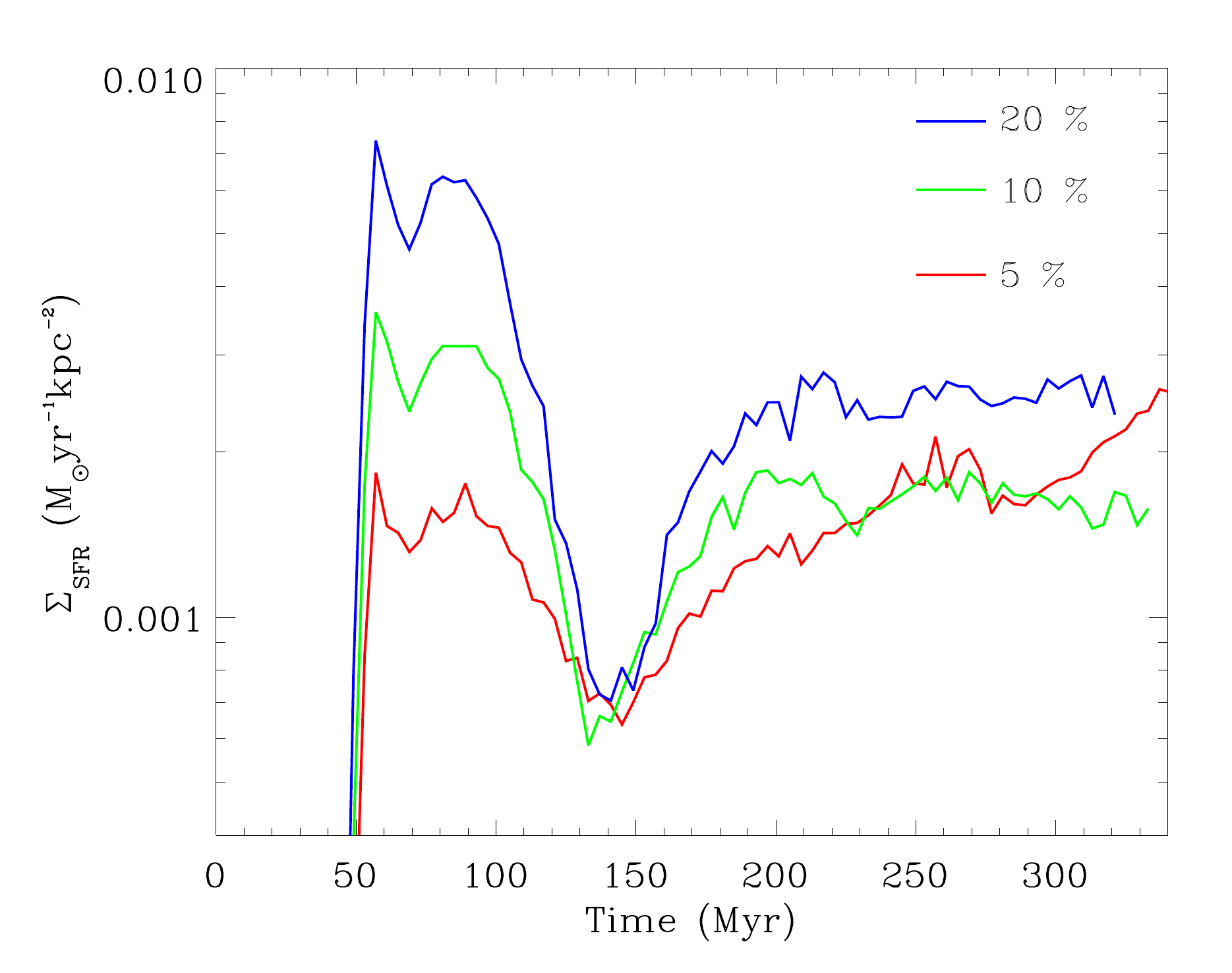}}
\caption{The star formation rates are shown here versus time for the
  calculations with different efficiencies (Runs L5, L10 and L20). The star formation rate is
not very steady, particularly at the early stages of the calculations. Simulations with
higher efficiencies generally have higher star formation rates, but by 200 Myr the
correspondence is less than linear, since with higher efficiencies
the gas has higher velocity dispersions and less is in the cold, dense
phase.}
\end{figure}
We show in Fig.~11 the star formation rate versus time for the
calculations with different efficiencies. The star formation rate
is not particularly steady for the duration of these calculations,
particularly for the higher efficiency cases where the star
formation rate is very high initially and then drops dramatically (see
also \citealt{Papadop2010}). As seen in Fig.~4, the gas in the disc
initially cools, which leads to a high level of star formation. This then
produces a large degree of warm gas, and correspondingly a large drop in
the rate of star formation. The
star formation rates then appear to
show less dramatic changes with time. 
One can see that during the
initial burst of star formation, the star formation rate scales with
the efficiency. However at later (after 150 Myr) times, the star formation rates are
much more similar regardless of the efficiency. This is
because after the initial phase of star formation, the models with
high efficiency have generated more hot gas and there is less material
available for star formation, whereas the models with low efficiency
have a higher fraction of cold, dense gas. The velocity dispersion of
the gas is also higher in the higher efficiency cases (Section~4.4).
So for example when $\epsilon= 20$ per cent, the star
formation rate is only around double that compared to when $\epsilon=
5$ per cent at 200 Myr. At later times, the star formation rate for the 5 per
cent efficiency case even exceeds that of the 10 per cent. As we will
discuss in Section 7, this is due to a number of long-lived bound
clouds (note that our calculations include short-lived unbound and
bound clouds, but the long-lived clouds are by necessity bound). 
Such clouds are present only in calculations with low star
formation efficiencies; with higher efficiencies the feedback is always
sufficent to disrupt the cloud, and clouds are short-lived. Though not
very numerous, these clouds have a disproportionate effect on the star
formation rate and account for the continued increase in the star formation rate past
300 Myr for the 5 per cent efficiency calculation.

We show how the star formation rate varies
with the global surface density, and how our results compare to the
Schmidt-Kennicutt relation in Section~6, where we present different
surface density calculations.

\section{Results - Calculations with a non-spiral potential}
In this section we investigate the importance of spiral density waves
for star formation, and the structure and properties of the ISM. The
calculations L5$_{nosp}$ and L10$_{nosp}$ are the same as the
previous calculations L5 and L10 except the spiral component of the
potential is not included. Thus they are very similar to calculations
by \citet{Tasker2009}, and \citet{Wada2000}, although the former did
not include feedback and the latter were 2D.
\begin{figure}
\centerline{
\includegraphics[scale=0.27]{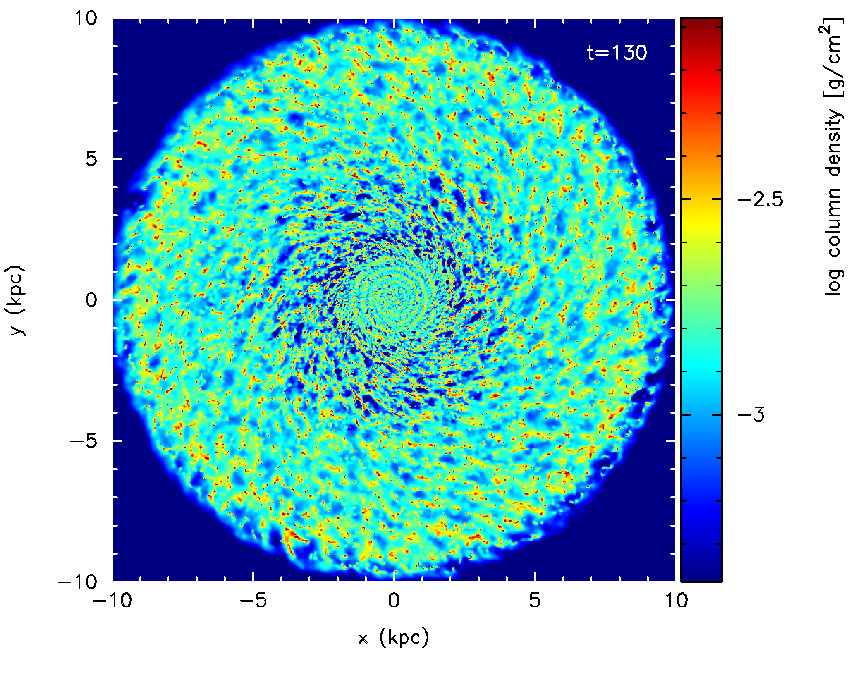}}
\centerline{
\includegraphics[scale=0.27]{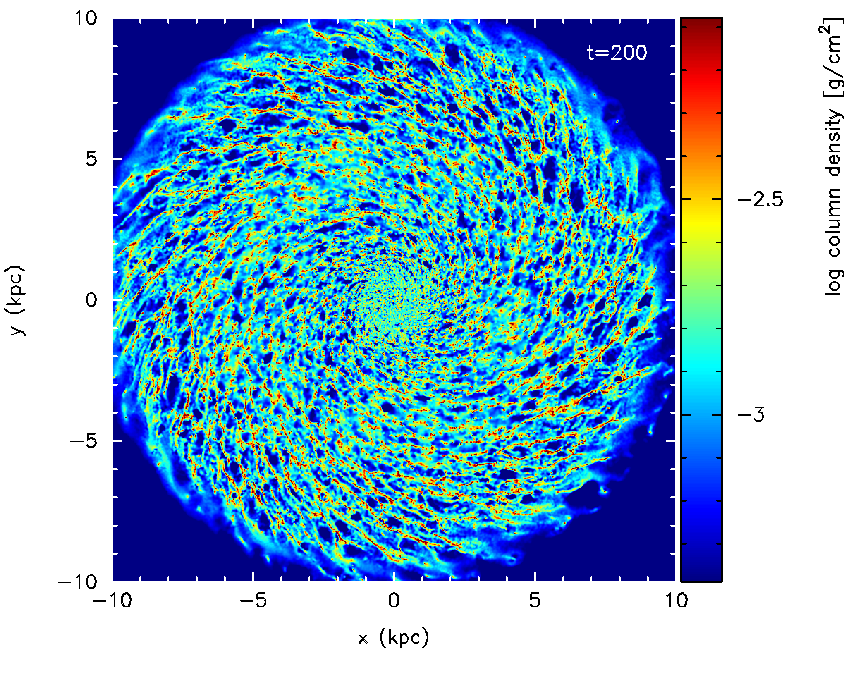}}
\caption{The column density is shown for Run L5$_{nosp}$, where
  $\epsilon=5$ per cent but there is no spiral component of the
  potential, at times of 130 and 200 Myr. At 130 Myr, the gas
  is much more diffuse (the amount of cold gas reaches a minimum at this point),
and appears less structured, following a peak in the star formation
rate. By 200 Myr, the structure is unchanging with time, and dominated
by the cold dense gas. The evolution of Run L10$_{nosp}$ is very
similar. Many holes can be seen at 200 Myr due to supernovae, which appear more irregular
and larger compared to those simply due to the flocculent structure.}
\end{figure}

We show the gas column density for the disc in Fig.~12. The gas is
highly structured, consisting of very many small spiral arm
segments. This is a somewhat simplified model since we
adopt a completely smooth disc potential. Very few flocculent galaxies
exhibit so much structure, though NGC~2841 is one possible example. 
Typically galaxies are susceptible to gravitational
perturbations in the stellar disc, which lead to long, multiple arms in the gas and stars 
simultaneously  (e.g. \citealt{Dobbs2007}).
Fig.~12 indicates however that the structure in the absence of a spiral
density wave is probably similar to the interarm structure in the
previous calculations with a spiral potential
(e.g.  Fig.~1). However the spiral arms in the models
shown in Fig.~1 gather the gas together, and determine
the spatial distribution of spurs.

\subsection{Properties of the ISM}
The scale heights, and amount of gas in the WNM tend to be slightly higher in Run
L5$_{nosp}$ compared to Run L5 (by $\sim$10--20 per cent), indicating the supernovae are having a
greater impact on the ISM. This is probably because the gas is not
gathered into spiral arms, so the hot gas can more readily diffuse,
and survive in the lower density regions. Otherwise, the distributions of the
cold, unstable and warm phases of the ISM in
the models without a spiral component are largely similar 
to those shown
earlier with a spiral density wave (Figs~4, 5, 6), and
likewise the scaleheights of supernovae events (Fig.~9). Hence we do
not show these results for the non-spiral models. The main difference in the calculations
with and without the spiral potential, is that with a spiral
potential, the gas is gathered into more massive clouds in the spiral
arms (see Section~7).

\subsection{Star formation rates}
\begin{figure}
\centerline{
\includegraphics[scale=0.42]{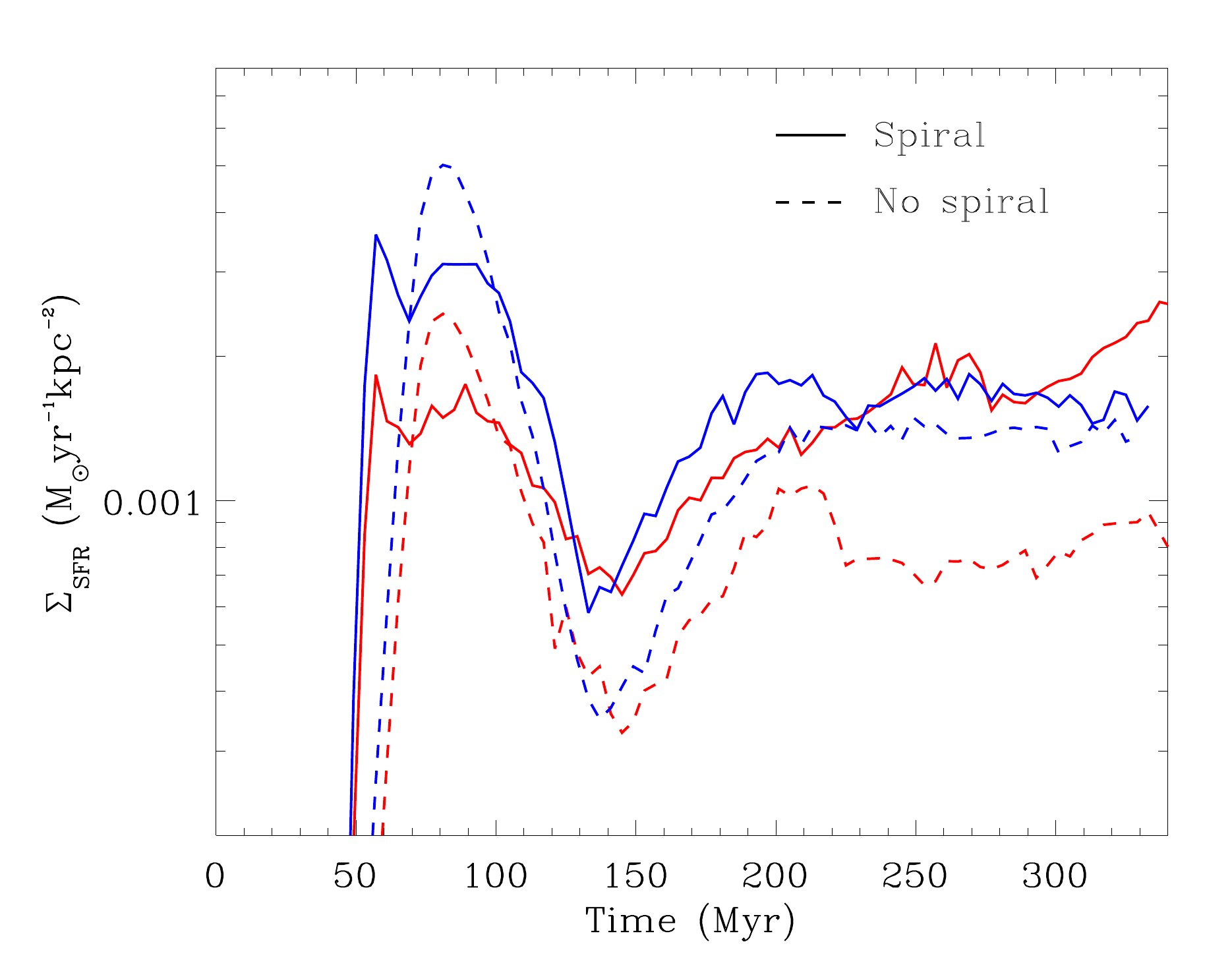}}
\caption{The star formation rate is plotted versus time for galaxies
  with and without a spiral potential. The red lines indicate the 5
  per cent efficiency calculations (Runs L5 and L5$_{nosp}$), whilst the
  blue lines indicate the 10
  per cent efficiency calculations (Runs L10 and L10$_{nosp}$).
 When $\epsilon=$5 per cent, the star formation is typically
  higher by factor of 2 or so with the spiral component of the
  potential. The difference is partly due to the presence of a few
  massive, bound, long-lived clouds in the case with a spiral
  potential. When $\epsilon=$10 per cent, by 200 Myr the star
  formation rates are similar, 
  regardless of the presence of the spiral component of the potential.}
\end{figure}
As discussed in the introduction, an important question is whether
spiral formation drives the triggering of star formation. 
In Fig.~13 we show the star formation rate for Runs L5, L5$_{nosp}$, L10
and L10$_{nosp}$
versus time, thus comparing the star formation rate with and without
the spiral component of the potential. For the 5 per cent efficiency
calculations (Runs L5 and L5$_{nosp}$), the star formation rate in the 
calculation with a spiral component is 2--2.5 times higher than
without. For the 10 per cent cases, the calculations evolve to
very similar star formation rates. With
the lower efficiency, the spiral arms are able to gather material into
more massive clouds and by the end of the calculation, there are
several long-lived, bound, massive clouds (Section 7). These clouds
have a disproportionate effect on the star formation rate. In contrast,
for the 10 per cent efficiency case, such clouds do not form
regardless of the presence of spiral shocks, so there is less
difference in the star formation rate. 

The presence and strength of the spiral potential also
determines when star formation begins in the calculation. There is
a higher initial star formation rate without a spiral potential -
possibly because star formation commences everywhere in the galaxy
simultaneously, rather than starting in the spiral arms.

\section{Results - Different surface density calculations}

\subsection{Structure and properties of the ISM}
In Fig.~14 we show the column density distribution for Run M10, which
has a higher surface density of 16 M$_{\odot}$ pc$^{-2}$, and an
efficiency of 10 per cent. The structure of the disc is fairly similar
to Run L10, with fairly continuous spiral arms, and clear spurs. The arms also
appear slightly wider than the lower surface density case.

For the calculation with an efficiency of only 5 per cent, we obtained
more massive clouds along the spiral arms. We include a section of the
disc, showing these massive clumps, in Fig.~15. Similarly to Run L1,
we cannot run this calculation further as the density of the gas
becomes very high in these massive, strongly bound and long--lived clouds. 
However we do note that the similarity between the structure along the
arm and simulations by \citet{Shetty2006}. In both cases,
the formation of the massive clouds is likely due to gravitational
instabilities rather than cloud collisions. Consequently the spiral
arms are not so continuous, the gas being instead arranged into
disproportionately massive clouds. In \citet{Shetty2006} these clouds
were stabilised by magnetic fields, which we do not include, and a
temperature floor of $10^4$ K. 

Unlike Run M5, for the case with a higher efficiency (M10), the feedback is
sufficient to prevent the formation of very strongly bound clouds,
thus the continuous gas spiral arms are maintained. Thus
perhaps rather counterintuitively the higher level of feedback helps
maintain the spiral structure in the gas.
\begin{figure}
\centerline{
\includegraphics[scale=0.27]{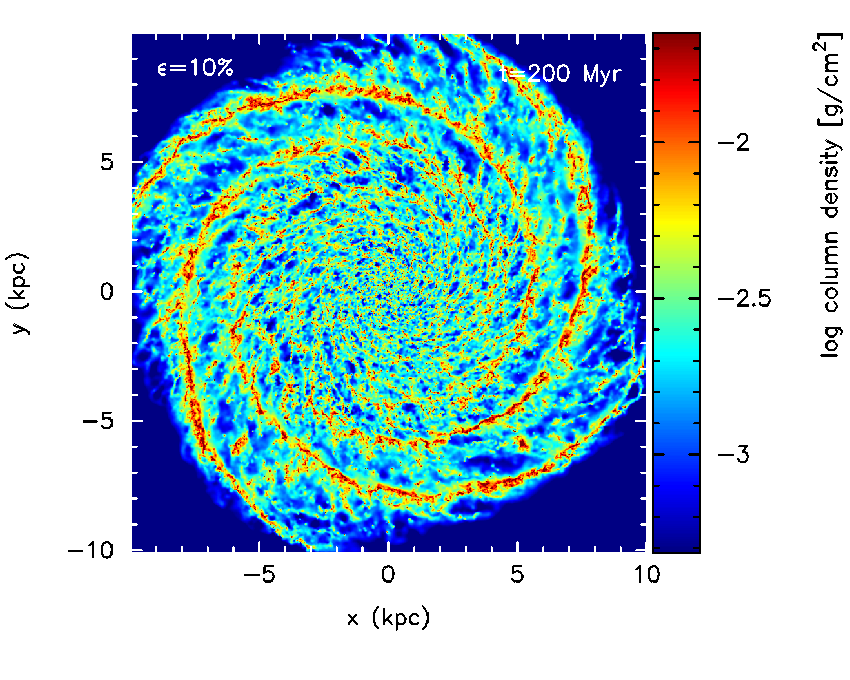}}
\caption{The column density is shown for Run M10, with a higher surface
  density of 16 M$_{\odot}$ pc$^{-2}$, and where
  $\epsilon=10$ per cent. The structure is fairly similar to the lower surface
  density model with $\epsilon=10$ per cent (Figure 1).}
\end{figure}

\begin{figure}
\centerline{
\includegraphics[scale=0.27]{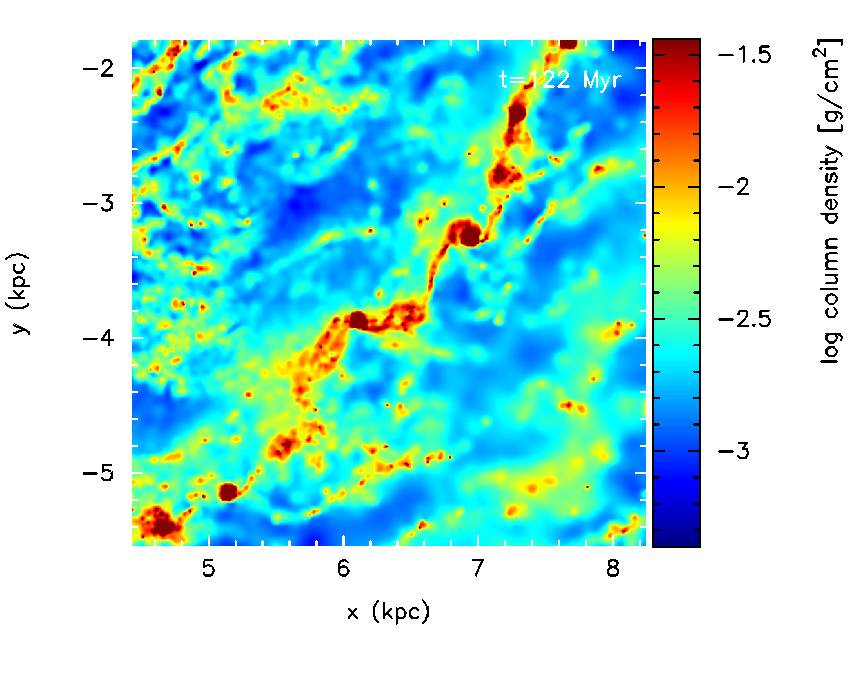}}
\caption{The column density is shown for a section of Run M5, at a
  time of 122 Myr. The figure shows the organisation of gas into
  dense discrete clumps along the arms, whilst continuous spiral arms cannot be maintained.}
\end{figure}

\begin{figure}
\centerline{
\includegraphics[scale=0.4]{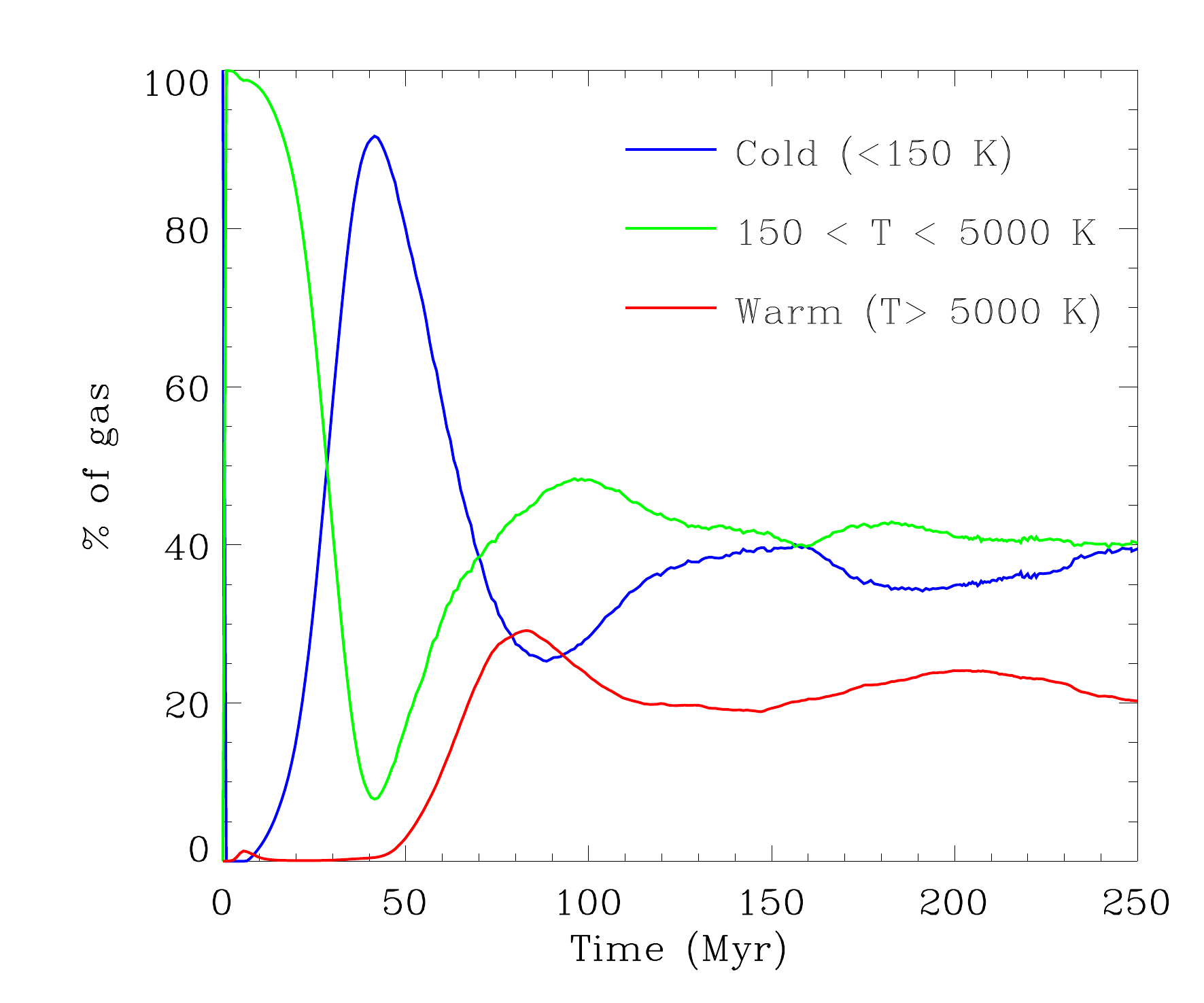}}
\caption{The fractions of gas (by mass) in the cold, unstable and warm
  regimes for the calculation with $\Sigma=16$ M$_{\odot}$ pc$^{-2}$
  and $\epsilon=5$ per cent (Run M10). The fractions of gas in the
  different phases is similar to those in the lower surface density
  models, but with slightly more cold gas.}
\end{figure}
In Fig.~16 we show the fraction of gas in different phases of the ISM
with time for Run M10, with $\Sigma=16$
M$_{\odot}$ pc$^{-2}$. The fractions of gas in the cold, unstable and
warm phases are similar to the lower surface density
calculations. There is slightly more cold gas than the equivalent 8 
M$_{\odot}$ pc$^{-2}$ calculation (Run L10), as expected for higher
densities. The scale heights, and velocity dispersions are slightly
higher (by around 25 per cent) in the intermediate compared to low
surface density calculations.
\subsection{Star formation rates}
\begin{figure}
\centerline{
\includegraphics[scale=0.4]{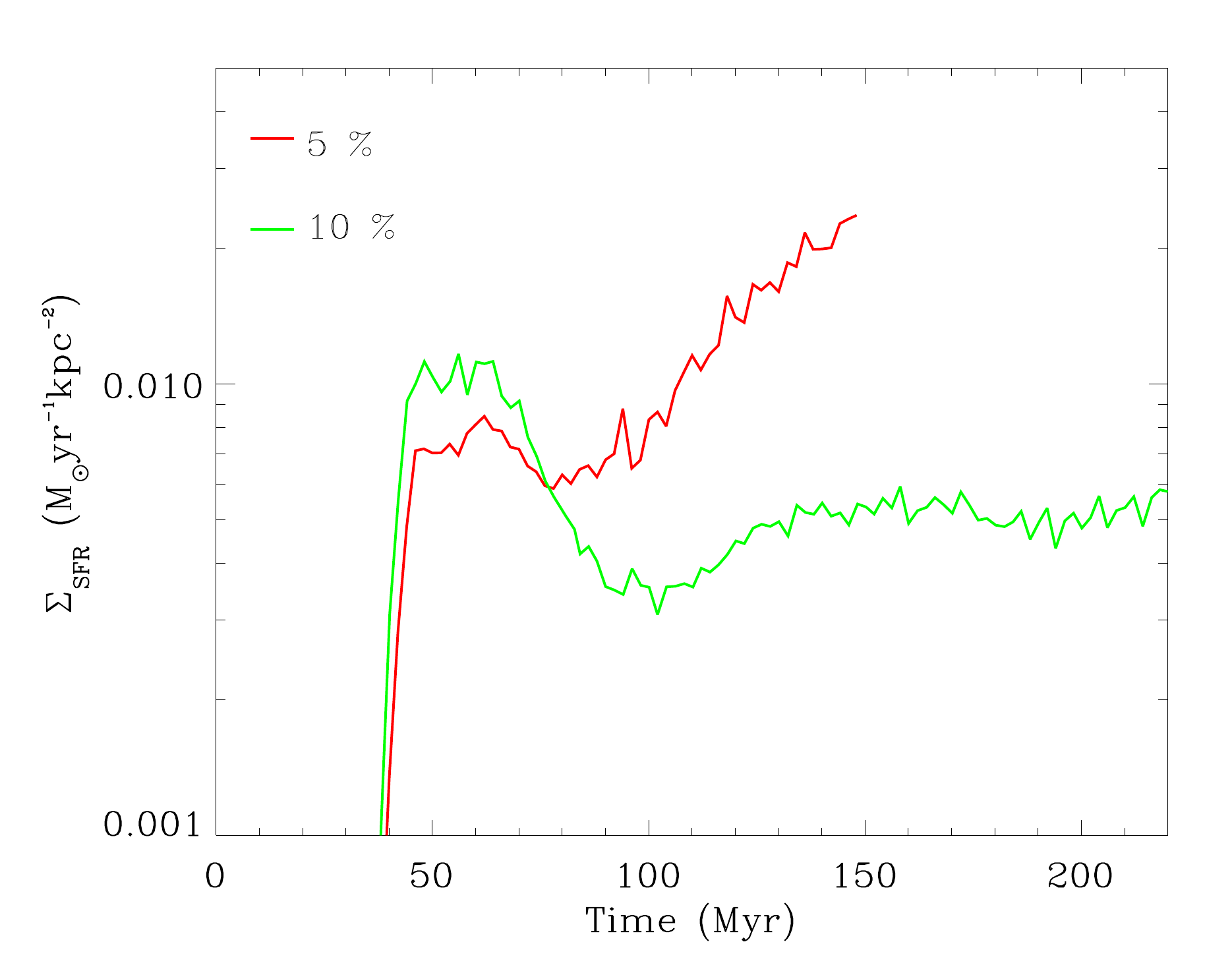}}
\caption{The star formation rates are shown from calculations M5 and
  M10, with a surface density of $\Sigma=16$ M$_{\odot}$
pc$^{-2}$. The star formation rate with a 5 per cent star formation
efficiency actually becomes higher than than with 10 per cent efficiency
because of the presence of long lived, massive, bound clouds, which
form stars continuously over a long time period.}
\end{figure}
In Fig.~17 we show the
star formation rate in Runs M5 and M10 (where $\Sigma=16$ M$_{\odot}$
pc$^{-2}$ and the star formation efficiency is 5 and 10 per cent respectively). For
the 10 per cent efficiency case, the evolution of the star formation rate is more
similar to the low surface density calculations; the star formation
rate is initially high, then decreases, and stays at a more or less
constant level. With 5 per cent efficiency,
the star formation rate increases instead, and is actually higher than for
the 10 per cent case. Again this is because in the 5 per cent
efficiency case, there are massive, bound, long lived clouds, which
have a marked effect on the star formation rate. 

The star
formation in Run M5 does not converge in our models. In reality, the lifetime of
clouds in Run M5 would be regulated by the conversion of gas to stars,
but we do not include this in our models. We found that the formation
of strongly bound clouds which are not disrupted, tends to occur at lower
efficiencies for higher surface density (40 M$_{\odot}$
pc$^{-2}$) calculations. Whilst this
trend is probably correct, we do not have the same mass resolution in the
high surface density calculations as the low surface density
calculations, so cannot make a true comparison.
\begin{figure}
\centerline{
\includegraphics[scale=0.6]{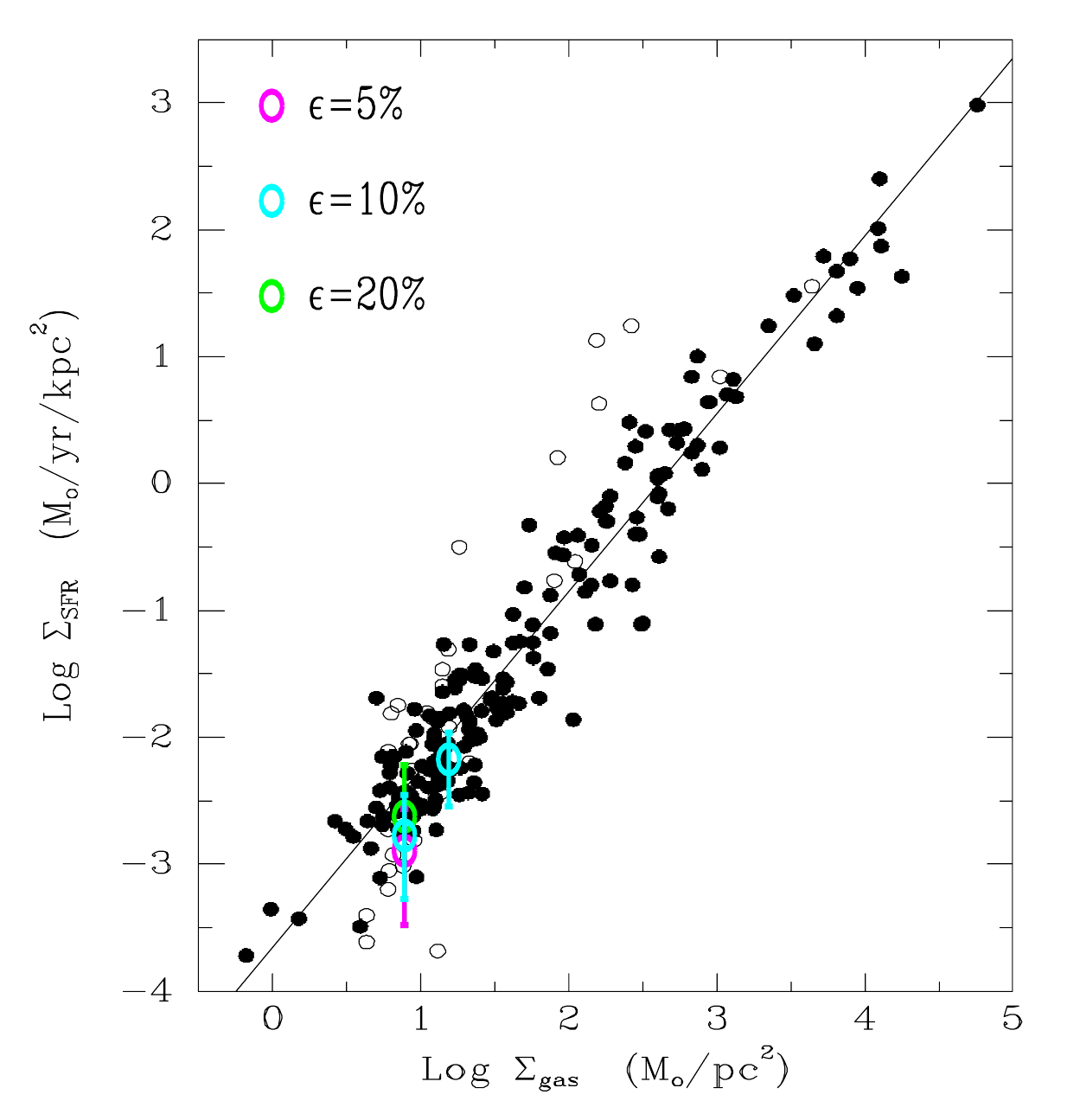}}
\caption{The star formation rates from the simulations with the spiral
  potential (Runs L5, L10, L20, M10) are plotted against surface density,
  the colours indicating the star formation efficiency. These points are
plotted on Figure 2 from \citet{Kennicutt2008}. Filled circles
denote luminous spirals an irregular galaxies with $M_B < −17$, and
open circles show fainter dwarf irregular galaxies. The error bars indicate
the range of the star formation rate over the duration the disc is
forming stars. 
The non-spiral calculations are
not included, but would lie roughly on top of the points shown here.}
\end{figure}

\subsection{The Schmidt-Kennicutt relation}
In this section we compare the star formation rate with the
  observed rates for galaxies, noting again that we do not include
  any a priori assumptions based on observations in order to
  determine the star formation rate. Observationally there is a clear trend,
  the Schmidt-Kennicutt
relation \citep{Schmidt1959,Kennicutt1998}, 
between the star formation rate and the gas surface density. 
In Fig.~18 we show the star formation rate versus the total gas surface
density, for Runs L5, L10, L20, and M10, where the star formation rate
has converged. All include the spiral component
of the potential. We can only estimate the star formation
rates from the calculations, as they tend to be highly time
dependent. So we show the star formation rate
at 200 Myr, whilst the error bars
indicate the maximum and minimum values of the star formation rate
over the duration of the simulations. 

Generally the star formation rates in our models provide
  reasonable agreement with observations. Although for the low surface
  density calculations, these points lie where the
Schmidt-Kennicutt relation starts to turn over and there is a large
degree of spread, so it is more difficult to compare with observations. We
see however that the efficiency
does not strongly affect the spread in the star formation
rate. 
Furthermore, though for clarity we do not show the calculations without a
spiral component, these points would lie on top of those already
plotted. Thus we obtain a greater degree of spread simply from the
time evolution of the galaxy, and whether the galaxy contains
long-lived clouds, than the efficiency or presence of an imposed spiral pattern.

\section{Properties of clouds}
In \citet{Dobbs2011} we focused on the virial parameters of the
molecular clouds formed in the simulations, and also discussed their
aspect ratios. We found that with a star
formation efficiency of 5 or 10 per cent, most of the clouds exhibited
$\alpha>1$. With an efficiency of 1 per
cent, most of the clouds had values of $\alpha <1$, and many had an aspect
ratio of $\sim1$, in disagreement with observations. 
In this section we give a broader perspective of the properties of the
clouds. We show properties from all the calculations except M5, which
does not show convergence. Although Run L1 also does not show
convergence, we retain this as a comparison to show cloud properties
when feedback is minimal. We select
clouds as described in \citet{Dobbs2011}.

\subsection{Masses and virial parameters of the clouds}
In Fig.~19 we plot the mass versus radius, and the distribution of
virial parameters for calculations L1 (after 125 Myr), L5, L10,
L5$_{nosp}$ and
M10 (at times of 200 Myr). These are the low surface density calculations with 1, 5 and 10
per cent star formation efficiencies, the $\Sigma=16$ M$_{\odot}$
pc$^{-2}$ calculation with $\epsilon=10$ per cent, and the simulation
with no spiral potential and $\epsilon=5$ per cent. The virial
parameters are calculated as described in
\citet{Dobbs2011}, where we aimed to match observational
determinations of $\alpha$ \footnote{We 
also calculated the virial parameter directly from
   the kinetic and potential energy of the clouds in model L5. These
    virial parameters are on average equal to, or higher by a
   factor of up to 2, than the $\alpha$'s shown in
   Figure~19.}. For model L5, we find one third of the clouds
   have $\alpha<1$ (by number, or 36\% by mass), 
one third are marginally bound with $1<\alpha<2$,
   and one third are unbound with $\alpha>2$. We note that we omit thermal and
   magnetic energy, which would increase $\alpha$.  We also did not
   include external pressure, though this does not readily lead to
   clouds in virial equilibrium \citep{Field2011}.

\begin{figure}
\centerline{
\includegraphics[scale=0.25, bb=0 0 530 350]{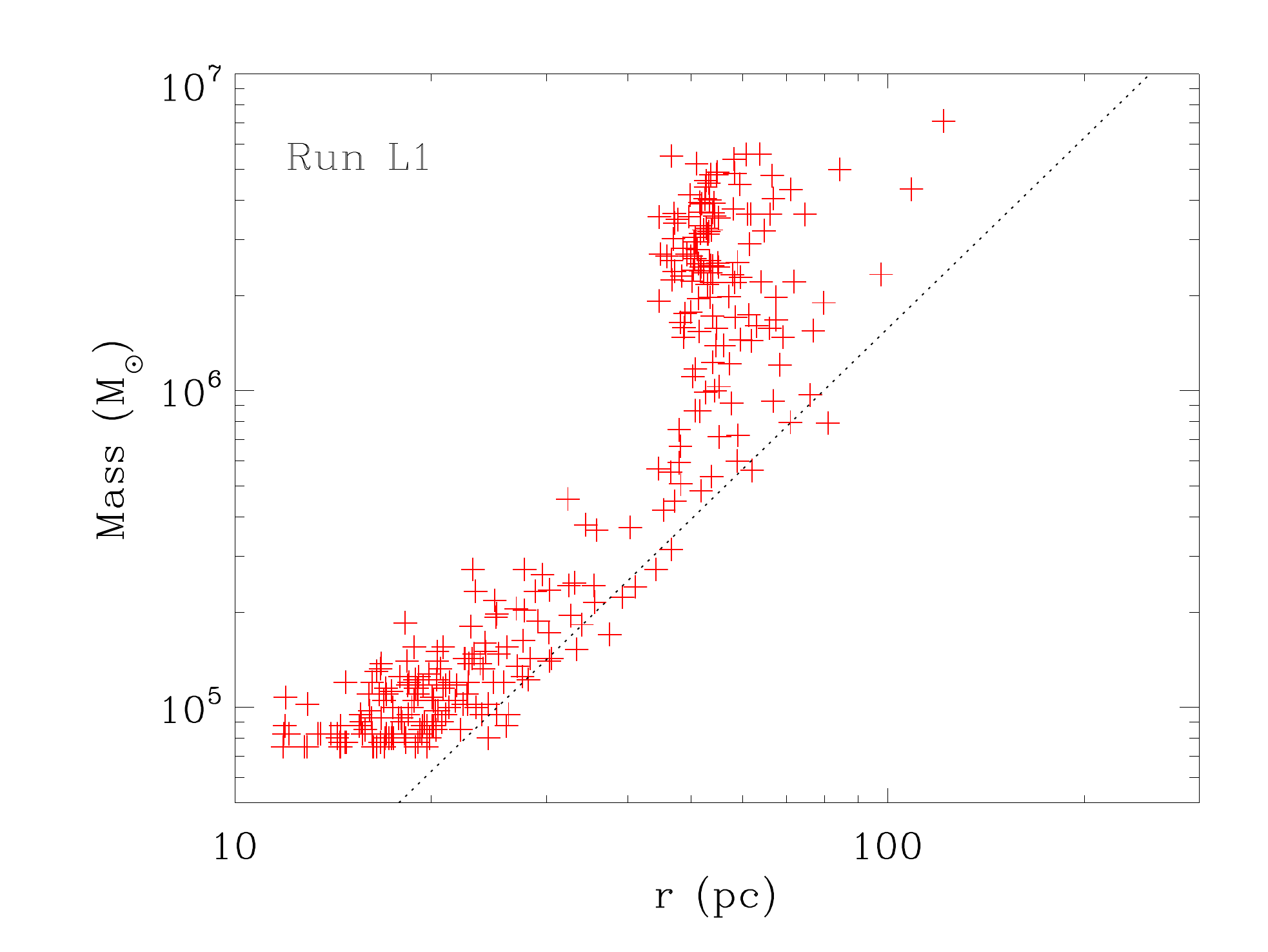}
\includegraphics[scale=0.25]{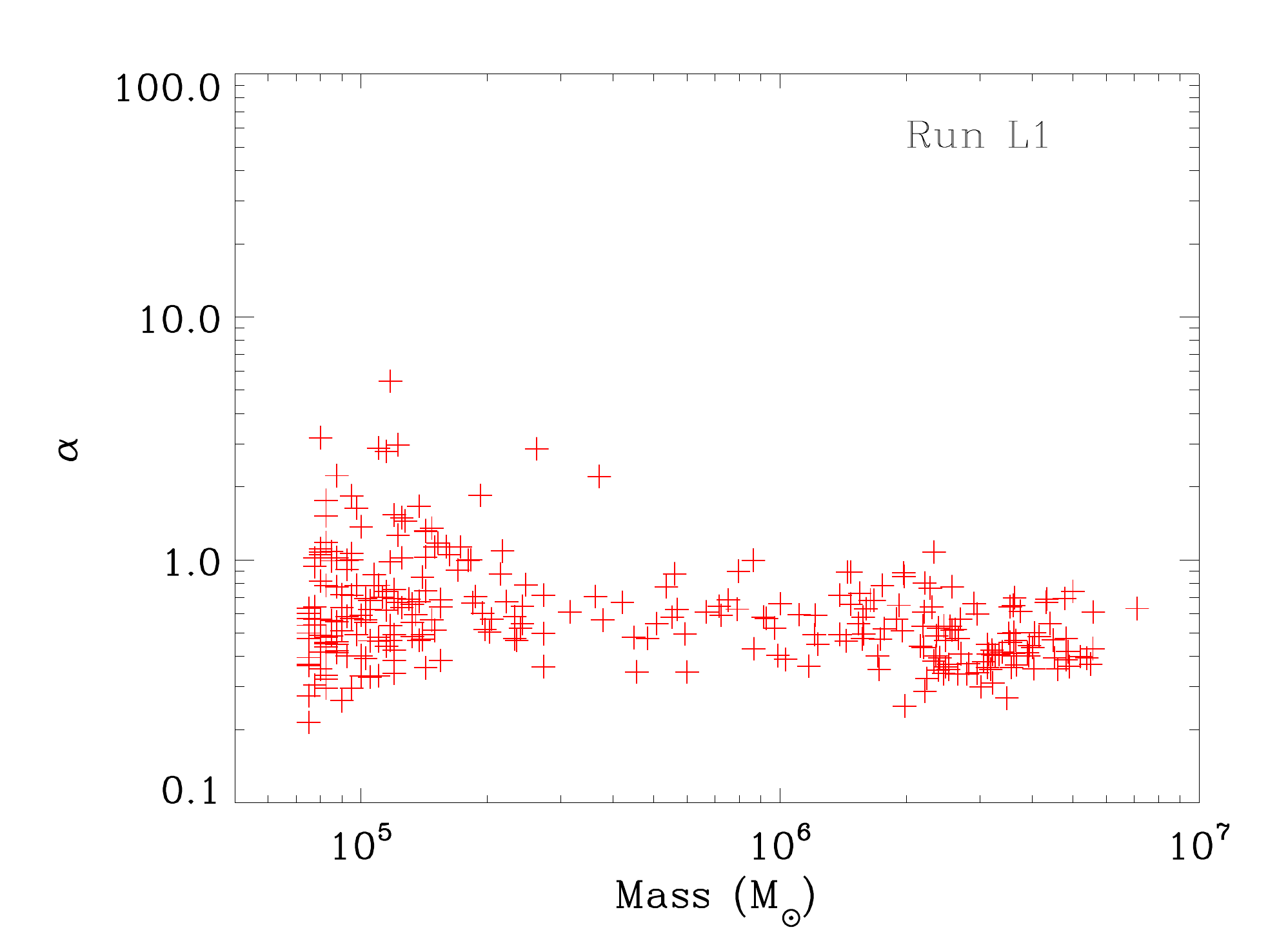}}
\centerline{
\includegraphics[scale=0.25, bb=0 0 530 350]{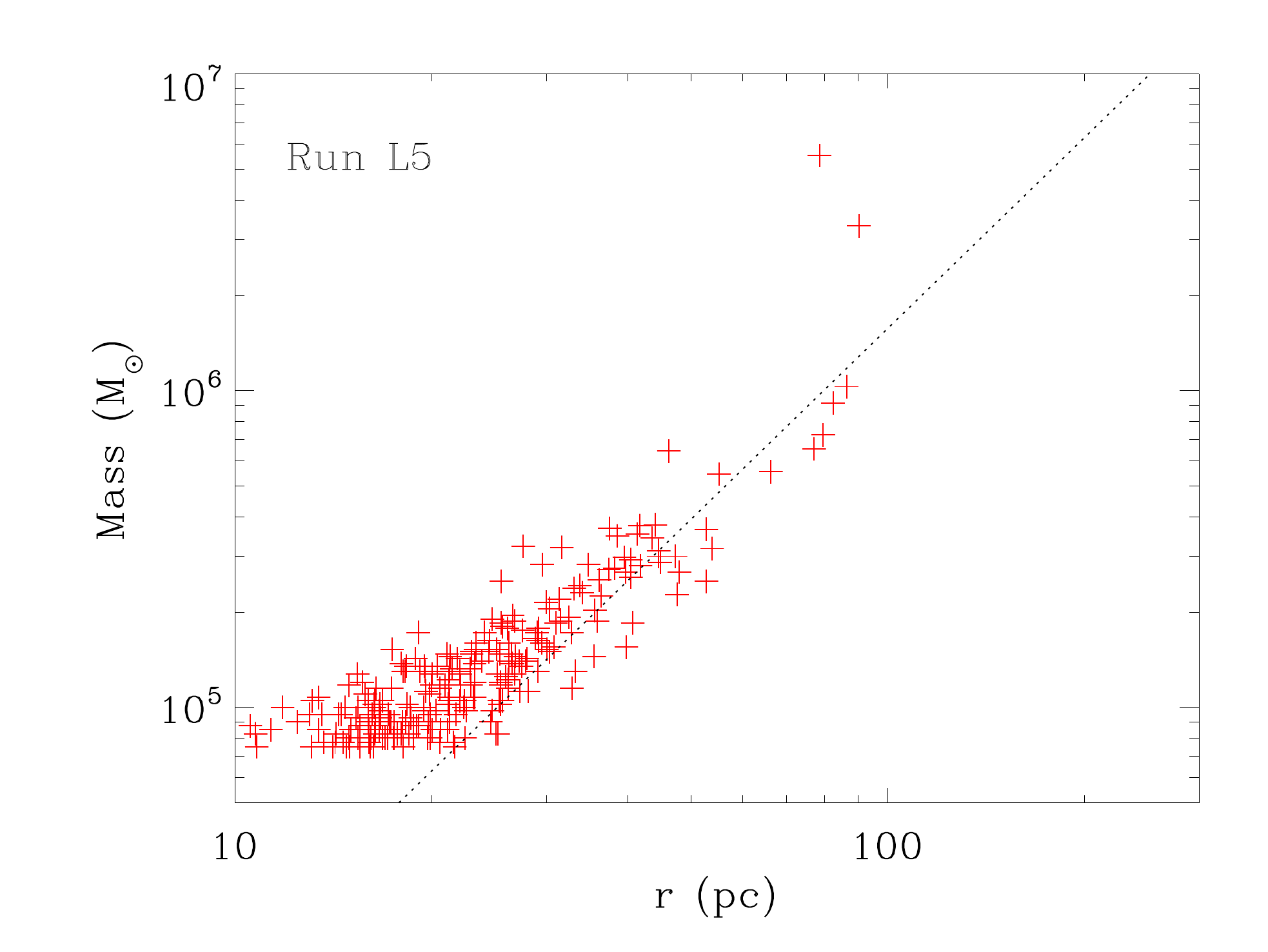}
\includegraphics[scale=0.25]{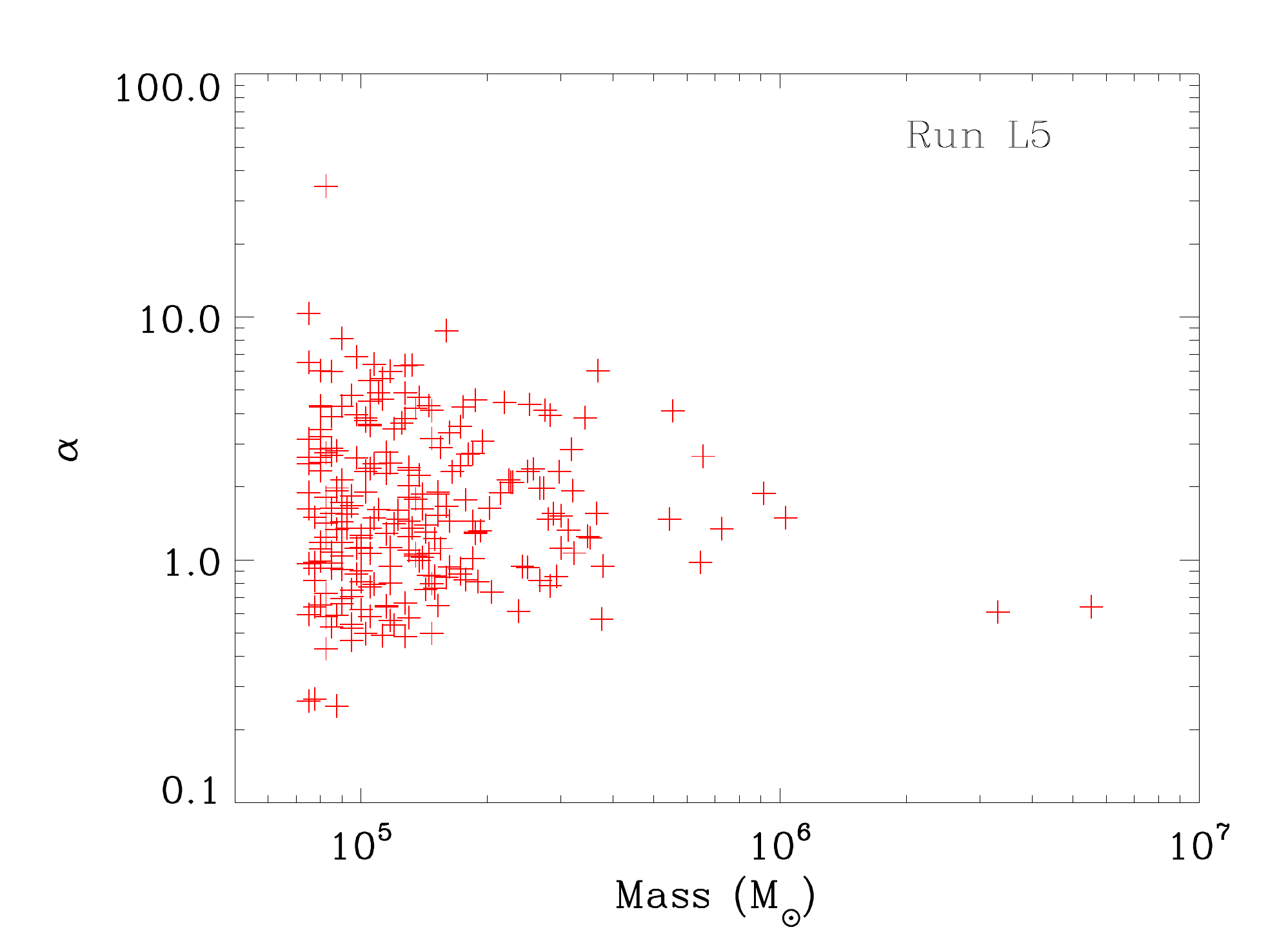}}
\centerline{
\includegraphics[scale=0.25, bb=0 0 530 350]{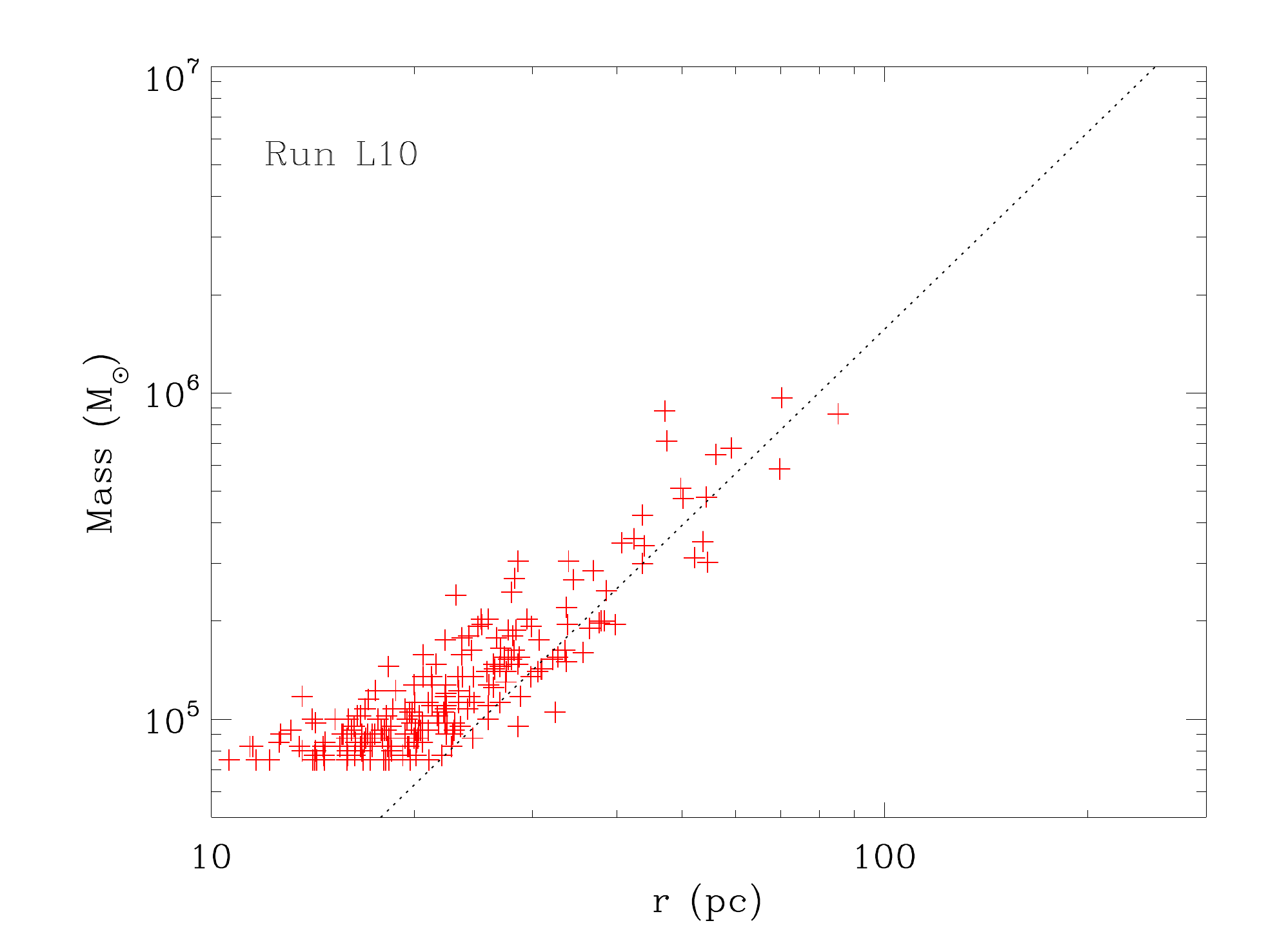}
\includegraphics[scale=0.25]{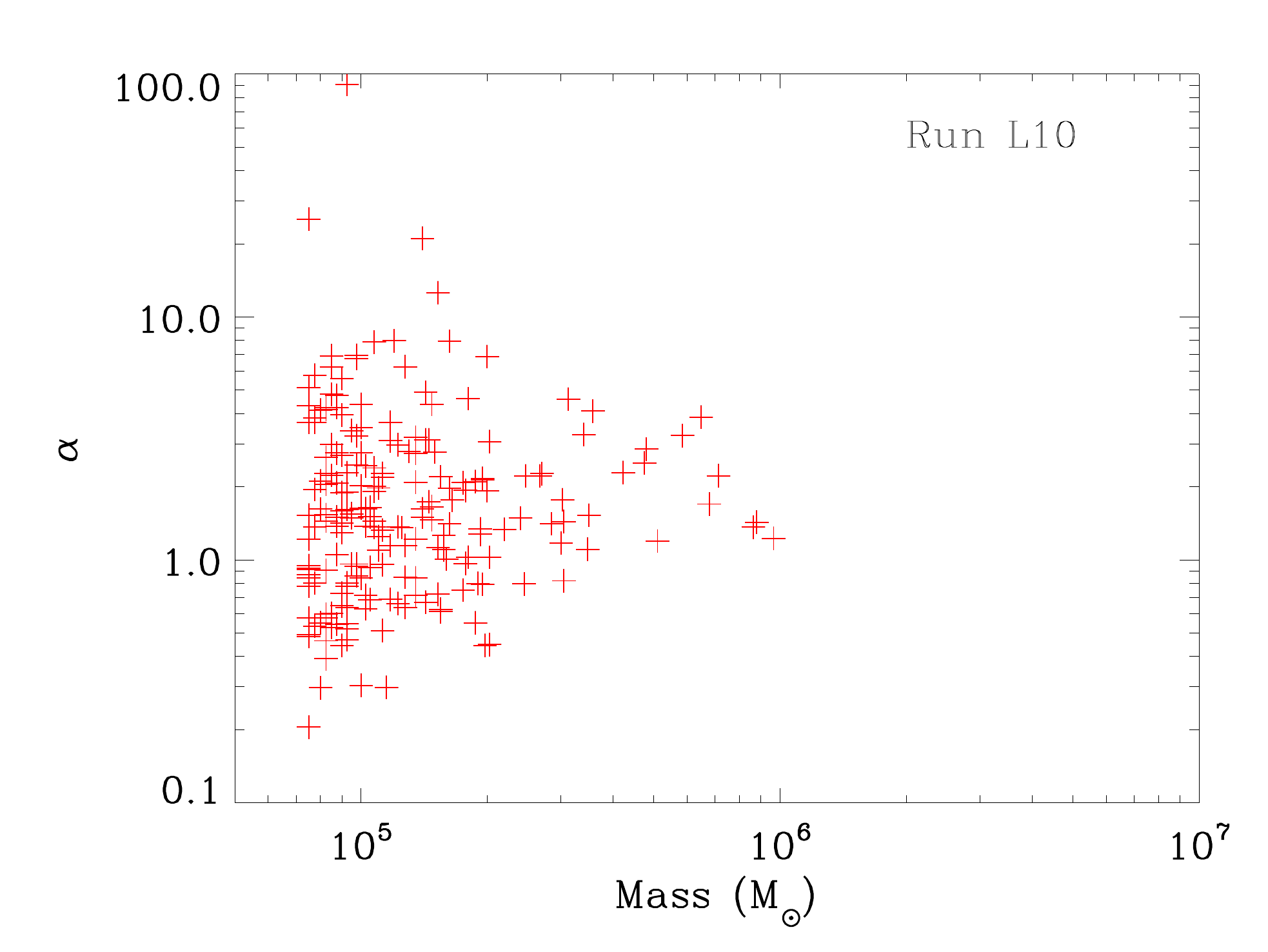}}
\centerline{
\includegraphics[scale=0.25, bb=0 0 530 35]{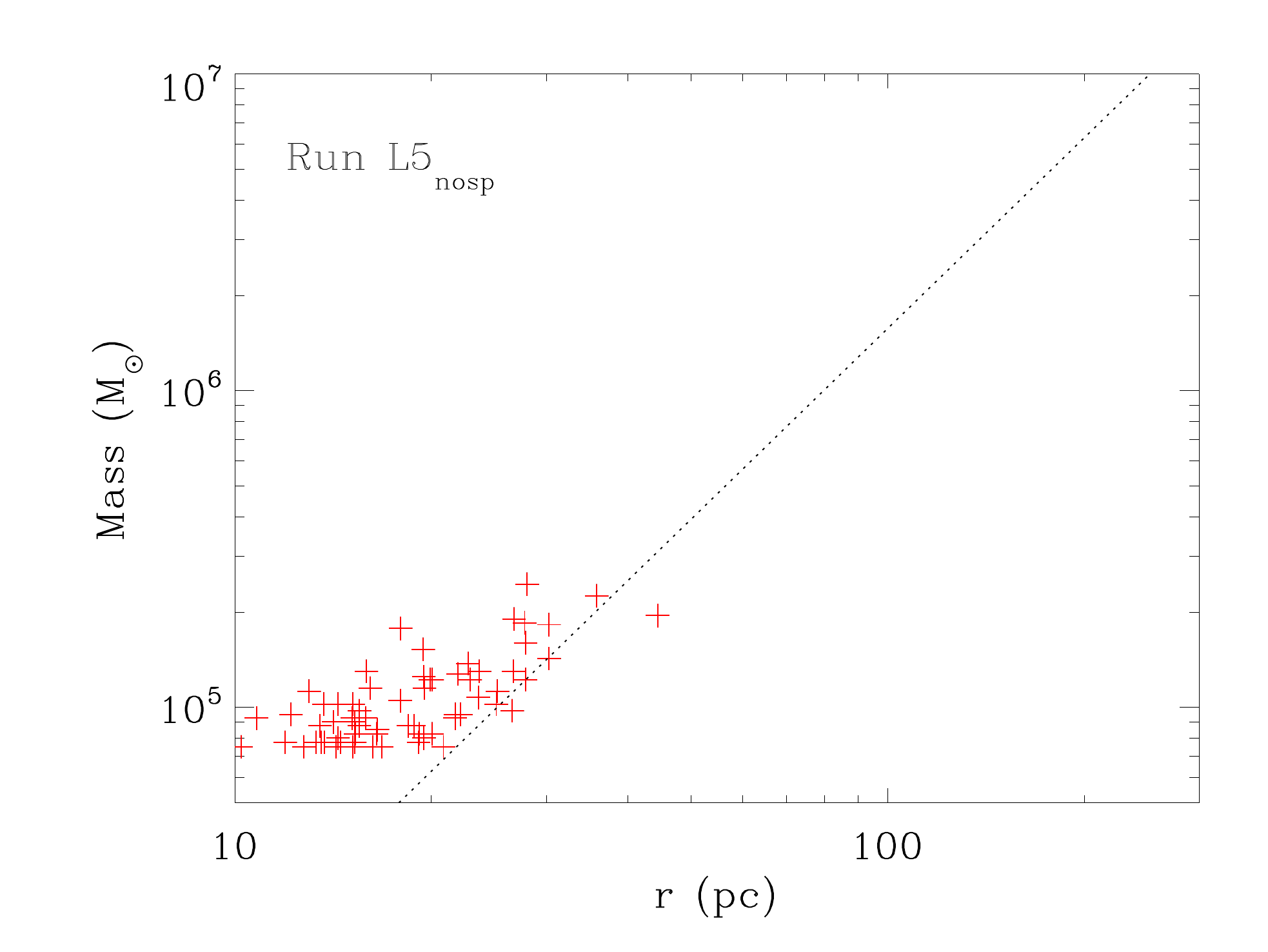}
\includegraphics[scale=0.25]{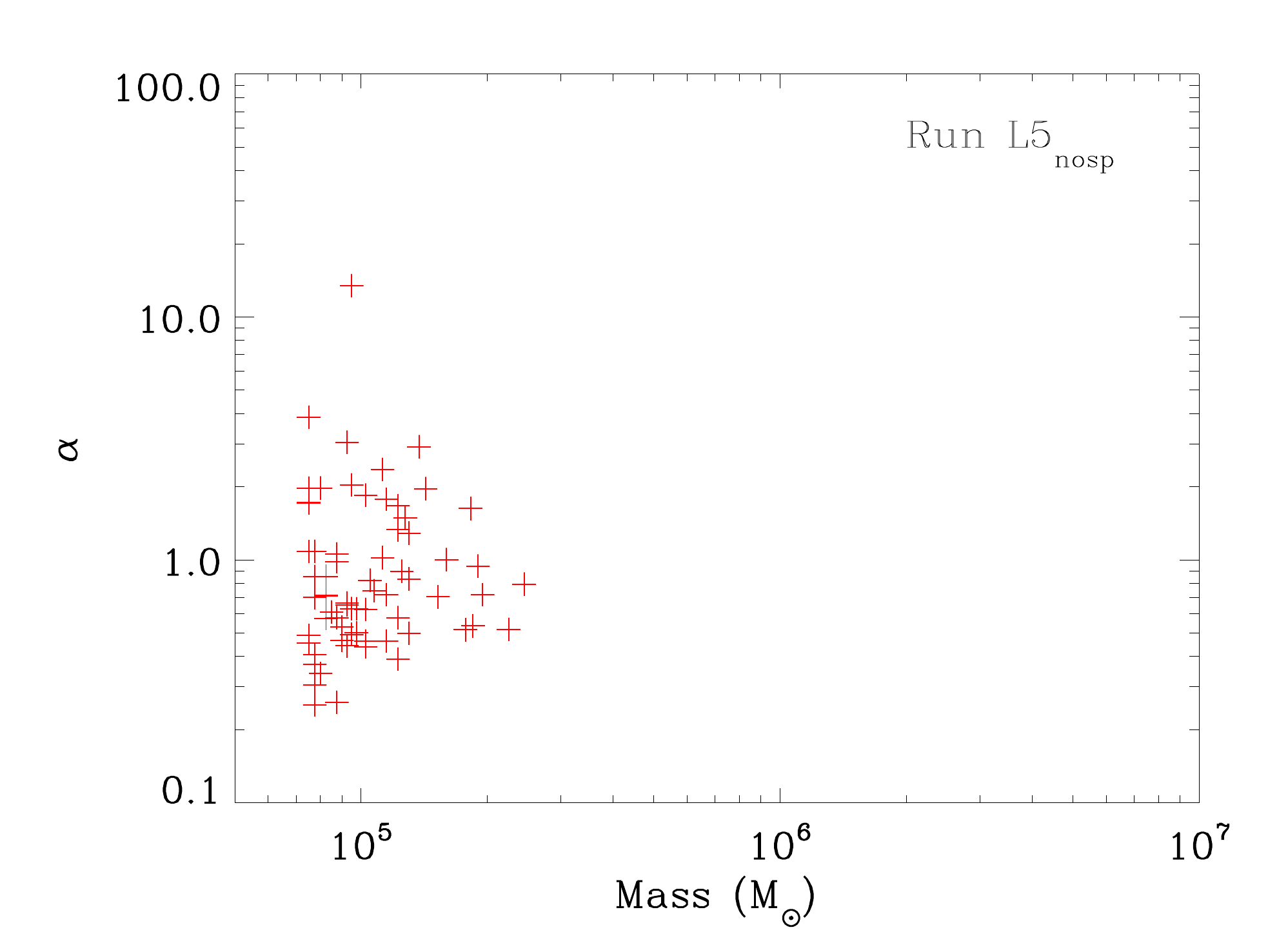}}
\centerline{
\includegraphics[scale=0.25, bb=0 0 530 35]{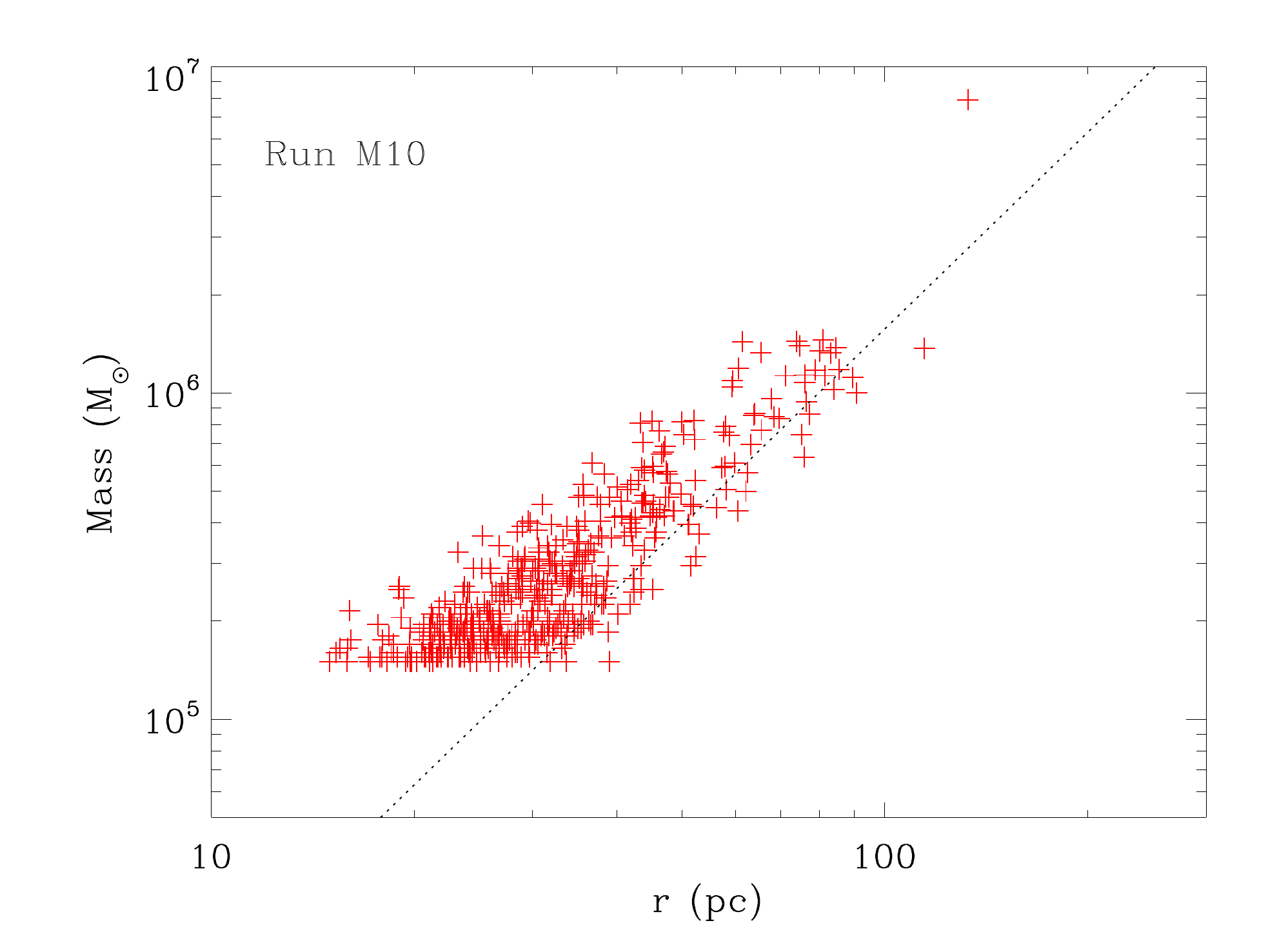}
\includegraphics[scale=0.25]{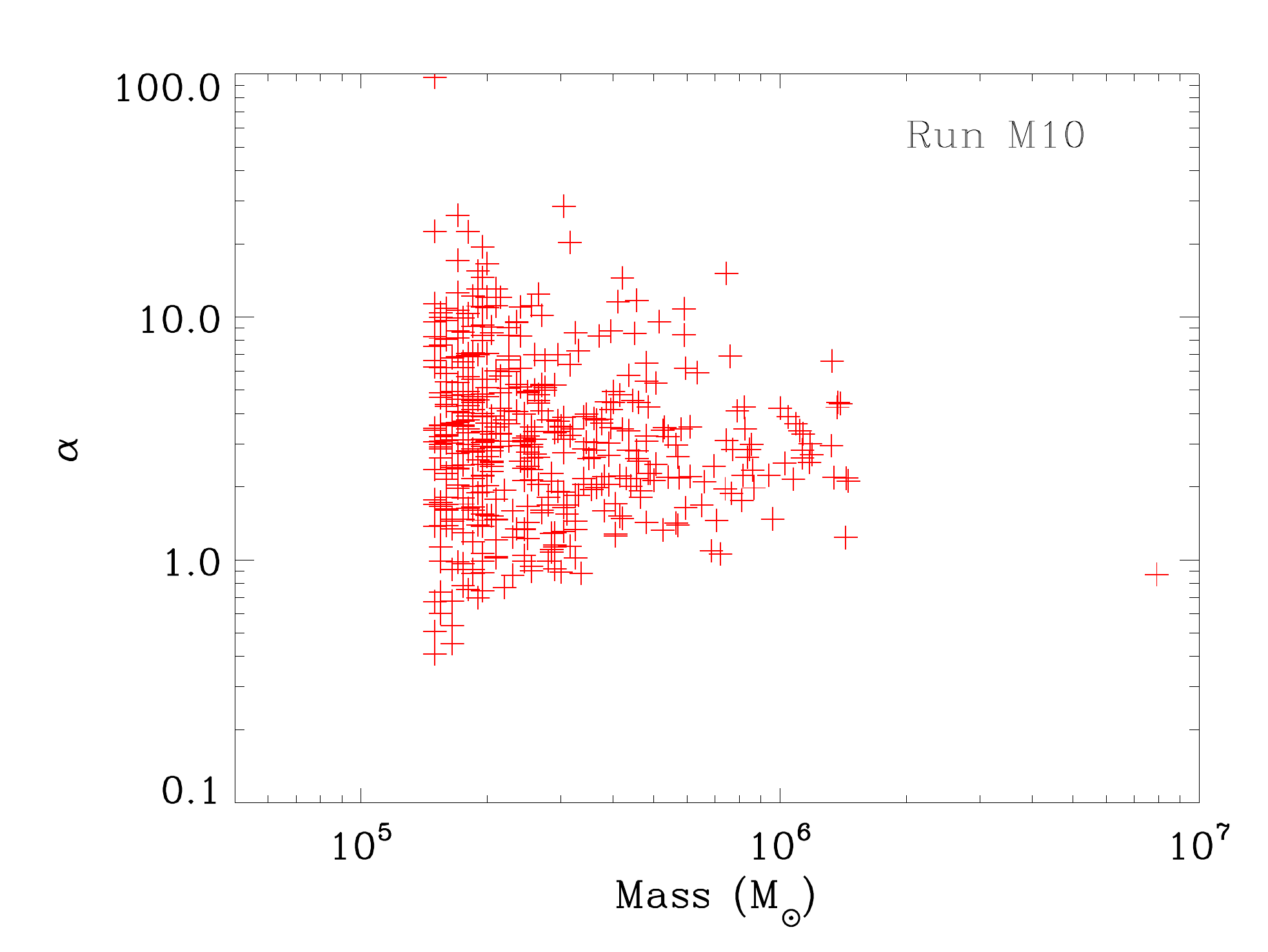}}
\caption{The masses and radii of the clouds are plotted (left), whilst their virial
  parameters are indicated on the right. The panels (in
  descending order) show the 8
  M$_{\odot}$ pc$^{-2}$ surface density calculations with 1, 5 and 10
  per cent efficiency, the calculation with no spiral potential and 5
  per cent efficiency, and the calculation with 16 M$_{\odot}$ pc$^{-2}$ and 10 per
  cent efficiency. The low efficiency cases, and higher surface
  density calculations exhibit clouds which are disproportionately
  massive, and strongly gravitationally bound. The dotted lines
  indicate a constant surface density of 50 M$_{\odot}$ pc$^{-2}$.}
\end{figure}

From Fig.~19, we see that the clouds in a given galaxy model
generally exhibit similar surface densities, with little spread in the
mass versus radius. The exception is the low efficiency model L1
($\epsilon=1$ per cent), where there are overly dense clouds. In this
case, where there is minimal feedback, the clouds can continue to
increase in mass without being disrupted. These clouds also exhibit
low virial parameters, and have long lifetimes. There are a couple of such 
massive clouds in Run L5. By 350 Myr in L5, these 2 clouds exceed
$10^7$ M${\odot}$ whilst several more long-lived bound, massive,
clouds have also formed. 

The departure of the clouds from constant surface densities is likely
related to the Jeans mass for a thin disc, $M_J=c_s^4/G^2\Sigma$,
where $c_s$ is the sound speed and $\Sigma$ is the surface
density. Taking instead of the sound speed, a velocity dispersion
$\sigma \sim7$ km s$^{-1}$ for model L5, and a surface density of 50
M$_{\odot}$ pc$^2$, gives $M_J=2.4 \times 10^6 $M$_{\odot}$. Thus
clouds above this mass are not supported by the velocity dispersion
against global gravitational perturbations. This is only an approximate
estimate, especially since the velocity dispersions are not equivalent to an isotropic
sound speed, however this value fits almost exactly with the turnover in
Figure~19 (second panel) The turnover for the 1 per cent
calculation also appears to be at lower masses which corresponds to
slightly lower velocity dispersions in that calculation. If we changed the
criteria for selecting clumps (and thereby $\Sigma$ above), we would
change the mass where this turnover occurs accordingly. 

The second and fourth panels in Fig.~19 compare calculations with and
without the spiral component of the potential (Runs L5 and
L5$_{nosp}$). There is a clear cut
off at $3 \times 10^5$ M$_{\odot}$ for the clouds from the calculation
with no spiral potential. Even for masses below $3 \times 10^5$ M$_{\odot}$,
there are significantly fewer clouds in the run with no spiral
potential. Thus the spiral density wave is vital for gathering gas
together to produce larger clouds. Though we have only a relatively
small sample of galaxies, observations also seem to indicate that
molecular clouds are less massive in galaxies without a strong spiral
pattern, e.g. M33 \citep{Rosolowsky2003}. For both the 10 and 20 per cent
efficiency cases, the masses of the clouds exceed those in
the calculation with no spiral potential.

Though we caution there are only a small number of clouds found in Run L5$_{nosp}$, and these
clouds are not well resolved, we found that there are a
higher fraction of bound clouds. With no spiral potential 70 per cent of the clouds are bound
compared to only 30 per cent with the spiral potential (even when just considering the lower
mass clouds). In other respects the clouds are similar. The
distributions of aspect ratios are similar, and the clouds in Run
L5$_{nosp}$ are short-lived. This difference in $\alpha$ could reflect
that the spiral potential allows gas to be gathered together
independently of gravity, whereas without the spiral potential, the
clouds are more often formed bound, undergo gravitational collapse, and are then dispersed. 

The third and fifth panels compare the properties of clouds in models
with different surface densities (L10 and M10, with 8 and 16
M$_{\odot}$ pc$^{-2}$ and
$\epsilon=10$ per cent). The properties of the clouds are
similar. There is a greater difference between models L5 and
M5 (as likewise between L1 and L5), but M5 does not show convergence.

We compute mass spectra for the calculations with different star
formation efficiencies in Fig.~20 (Runs L1, L5 and L20).
The cloud
mass function has a slope of around $dN/dM \propto M^{-1.9 \pm 0.1}$
for the clouds in Runs L5 and L20, with 5 and 20 per cent efficiencies
(and though not shown, the slope is similar for Run L10 with 10 per cent
efficiency). This is similar to the results we obtained in
\citet{Dobbs2008}, and to observations \citep{Heyer2001}. Thus the
star formation efficiency does not change the slope, rather the level
of stellar feedback determines the normalisation of the cloud mass
function. For the calculation with 20 per cent efficiency, there are
fewer clouds, and the maximum mass is $7.5 \times 10^5$ M$_{\odot}$.
Though we do not show it in Fig.~20, the cloud mass spectrum for Run M10,
with a higher surface density, has a slightly shallower slope 
($dN/dM \propto M^{-1.65 \pm 0.1}$), which could be due to the increased
importance of self gravity \citep{Dobbs2008,Dib2008}. The normalisation is also naturally
higher. 

The mass function for Run L1, with $\epsilon=1$ per cent is very
different. There is a peak at $5 \times 10^6$ M$_{\odot}$, which is
not seen in observations. Again these high mass clouds are gravitationally
bound, long lived clouds. In Run L1, feedback is insufficient to
disrupt the clouds, therefore they can continue to increase in
size. Their mass is instead only limited by their age, and the
accretion of gas from the surrounding medium. In fact any of the
simulations where we find a significant number of clouds which do not
follow a constant surface density relation (i.e. L1, eventually L5,
and M5) would exhibit such a
bimodal distribution.

\begin{figure}
\centerline{
\includegraphics[scale=0.45]{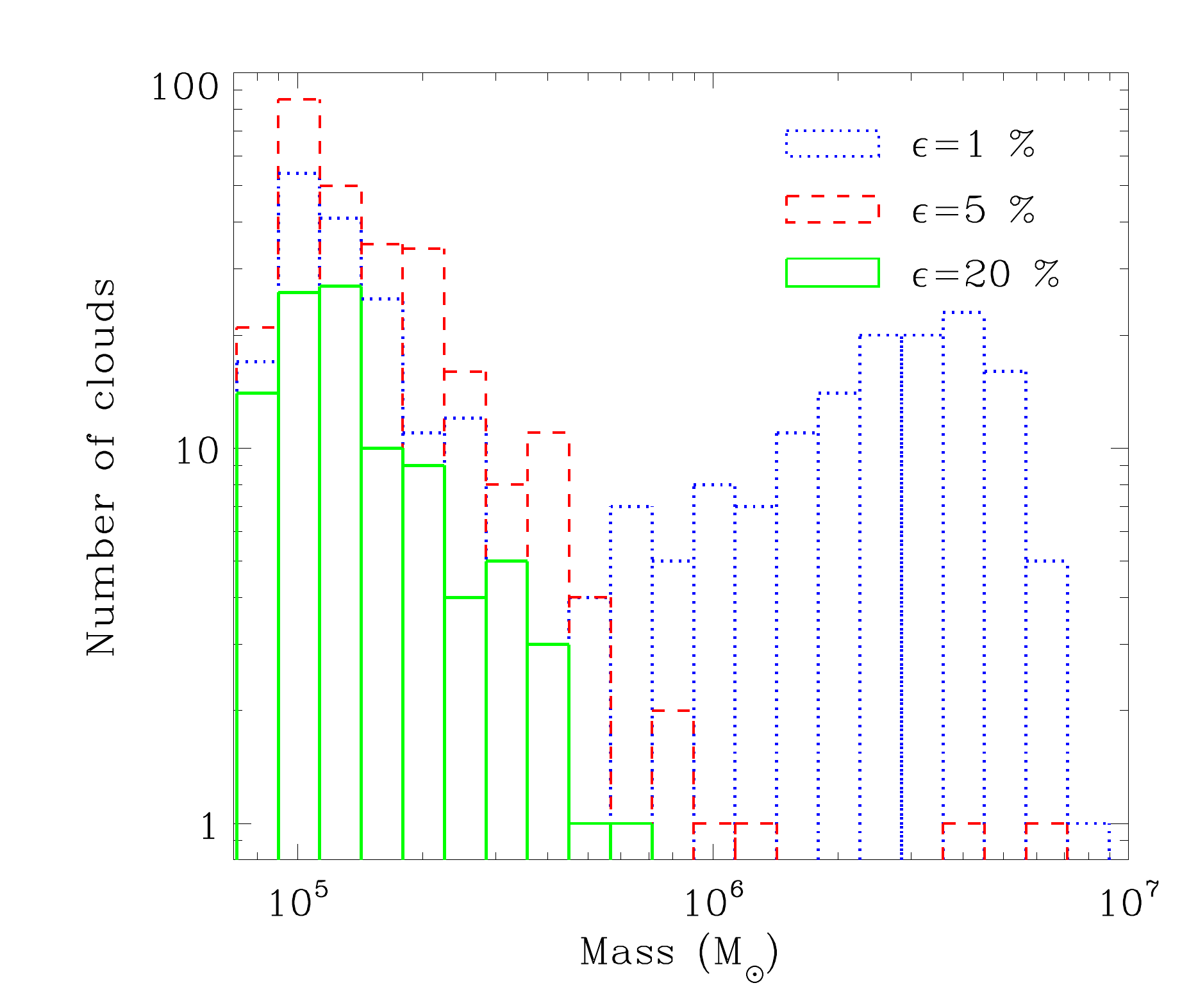}}
\caption{The mass spectra are shown for the calculations with
  $\epsilon =1, 5$ and 20 per cent (L1, L5 and L20), where there is a
  spiral potential. The spectra are calculated at a time of 200 Myr
  for Runs A5 and A20, and 125 Myr for Run A1. The slope of the mass spectra is about $dN/dM
  \propto M^{-1.9 \pm 0.1}$ when $\epsilon =5$ and 20 per cent, the
  mass spectrum is merely
shifted to lower masses with the higher star formation efficiency. The
spectra for the 1 per cent efficiency case (Run L1) 
exhibits a bimodal distribution due to a population of long-lived bound
clouds.}
\end{figure}

\subsection{Cloud heights and Orion}
Earlier in Fig.~9 we showed the heights of supernovae events in the
disc. We also determined the height of the clouds found in 
calculations L5, L10 and L20, at a time of 200 Myr. The maximum
heights of the clouds are $\sim$ 120, 220 and 300 pc for the
calculations with 5, 10 and 20 per cent efficiency respectively. Orion
lies at a height of 200 pc above the midplane. For the 10 and 20 per
cent efficiency calculations, clouds with heights similar to Orion are
relatively easy to produce. For the 5 per cent calculation, they are
much rarer. Fig.~9 shows that a small number of supernovae have occured at heights
of 200 pc, suggesting we would only find such a high latitude
cloud at particular time intervals. The scale heights are
  similar or higher for the
  higher surface density calculations.

\subsection{Cloud rotation}
The angular momenta of the clouds is shown in Fig.~21 for
Run L5. As
pointed out in \citet{Dobbs2008}, cloud collisions can be sufficiently
disruptive to cause GMCs to have a net rotation in a direction
opposite to the rotation of the galaxy. The
distribution is similar to that shown in \citet{Dobbs2008}, with
a similar fraction (37 per cent) of retrograde rotating clouds. The
low mass clouds have slightly higher angular momenta compared to
\citet{Dobbs2008}.
The distribution of angular momenta is similar for Run M10,
  though with a few more higher momentum clouds.
\begin{figure}
\centerline{
\includegraphics[scale=0.4]{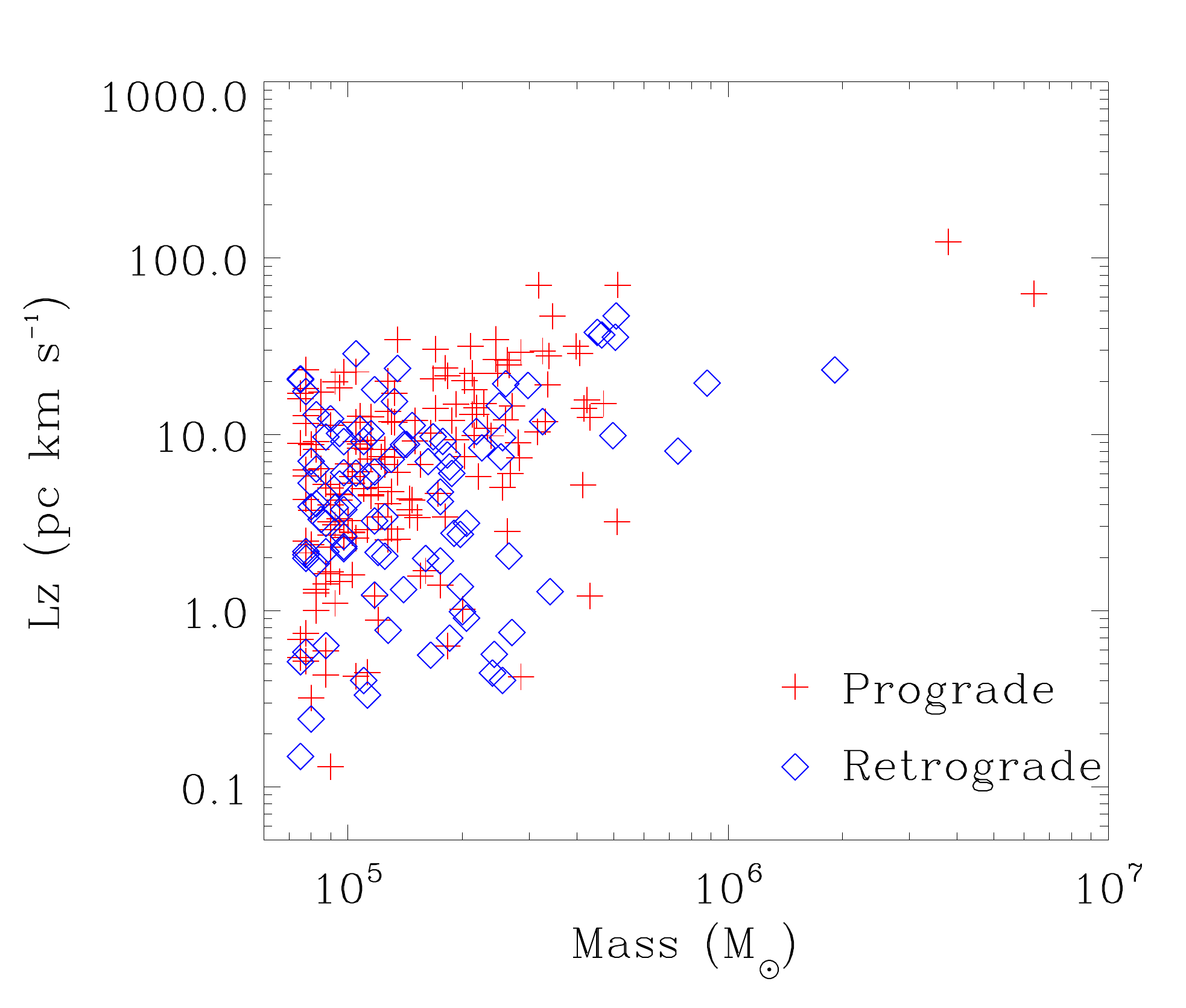}}
\caption{The angular momenta are shown for clouds from Run L5
  ($\epsilon=5$ per cent) at a time of 200 Myr. 
  Nearly 40 per cent of the clouds exhibit retrograde
  rotation, though the 2 very massive clouds, long-lived clouds are prograde.}
\end{figure}

We show the fraction of retrograde clouds for the different
calculations in Fig.~22. Observations of clouds in our Galaxy, and
M33, suggest that $\sim40$--60 per cent of clouds exhibit retrograde
rotation \citep{Phillips1999,Rosolowsky2003,Imara2011,Imara2011a}. 
For the calculations with $\epsilon=$
5, 10 and 20 per cent (L5, L10 and L20), and the medium surface
density calculation with $\epsilon=$ 10 per cent (M10), the fraction of retrograde
clouds agrees with observations. For Run L1 ($\epsilon=1$ per cent)
however, there are only $\sim 13$ per cent retrograde clouds, much
lower then observations. In this calculation, much of the gas accumulates into
$>10^6$M$_{\odot}$ clouds which rarely collide and are not
substantially disrupted by stellar feedback. Thus there is no
mechanism to cause these clouds to rotate retrogradely. Instead the
clouds continue to accrete gas due to self gravity.

We also show in Fig.~22 the fraction of retrograde clouds for Runs
L5$_{nosp}$ and L10$_{nosp}$. In both cases the fraction of retrograde clouds
is less than with the spiral potential. This again points to a
scenario where collisions are less important in the calculations
without a spiral potential, and the clouds mainly form by self
gravity.   
\begin{figure}
\centerline{
\includegraphics[scale=0.6]{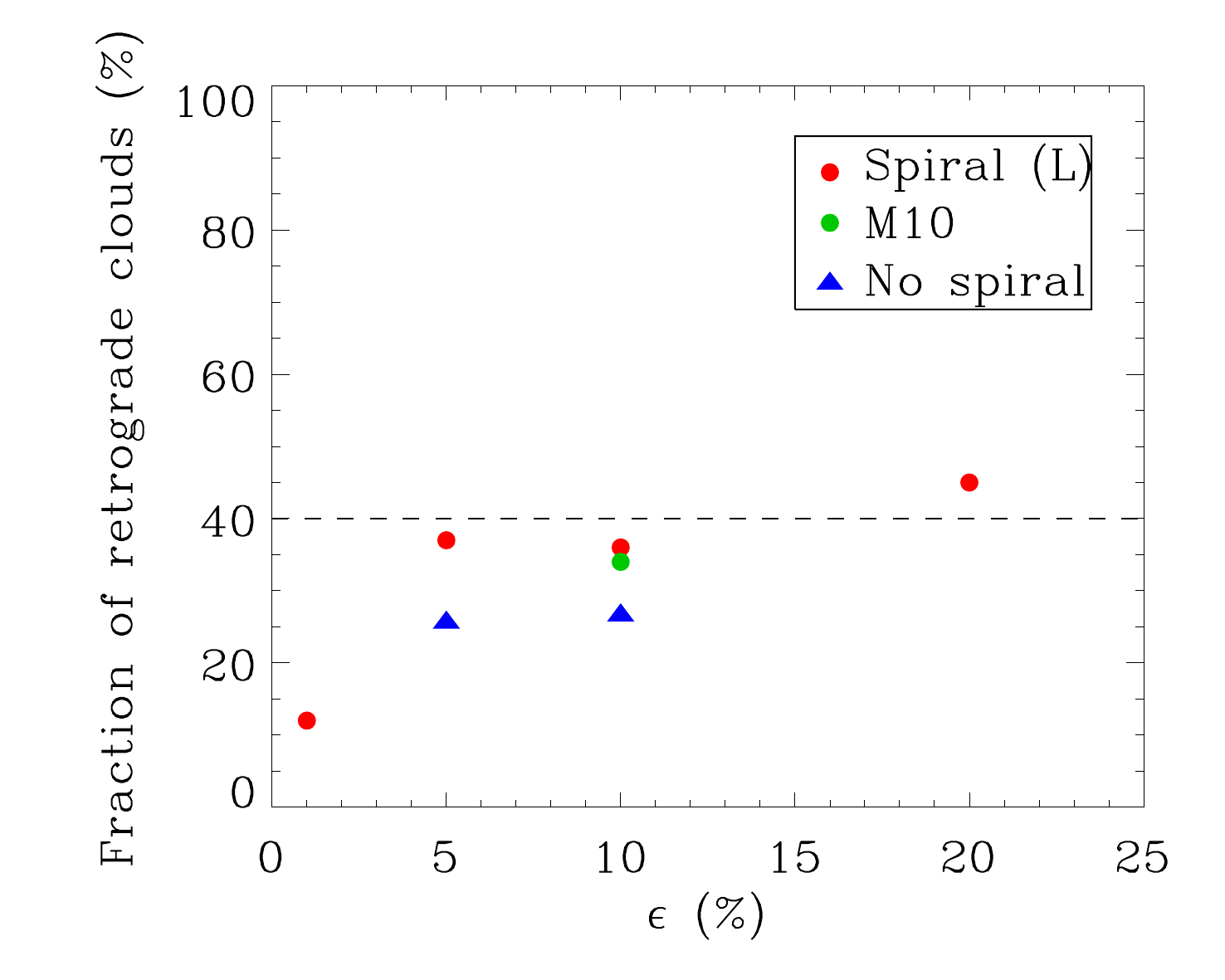}}
\caption{The fraction of retrograde clouds is shown versus star
  formation efficiency for the models L1, L5, L10, L20, M10, L5$_{nosp}$, and
  L10$_{nosp}$. The dashed line shows the fraction of retrograde clouds
  estimated for M33 \citep{Rosolowsky2003}. The fraction of retrograde
  clouds matches observations for the spiral calculations, with the
  exception of the $\epsilon=1$ per cent case, whilst the fraction is
  a little low without a spiral potential.}
\end{figure}
 
\section{Conclusions}
In order to model the ISM in spiral galaxies we have carried out
numerical three-dimensional hydrodynamic simulations of gas 
flowing in a fixed, full disc, global potential. The model involves 
a standard cooling law, and heating from background UV radiation 
together with localised input from star-formation episodes. 
Star formation is assumed to occur instantaneously when a parcel 
of the ISM becomes sufficiently compact and self-gravitating that it
undergoes dynamical collapse. When this occurs a fraction, $\epsilon$, 
of this gas is assumed to form stars and to provide instantaneous 
feedback to the ISM. Using these simple ideas we are able to reproduce 
the following features of the ISM:

\subsection{Gas properties}
 \begin{enumerate}
\item{We obtain roughly equal fractions of gas which are in the cold ($T < 150K$),
    intermediate, thermally unstable ($150K < T < 5000K$) and warm ($T >
    5000K$) phases. The relative fractions depend mainly on the assumed
    efficiency $\epsilon$ of star- formation feedback (Figure 4).}
\item{The scaleheights of the cold, unstable and warm phases are in
    reasonable agreement with observations. Here
    values of $\epsilon$ in the range 0.1 -- 0.2 give the best fit to
    the observations of the Galaxy (Figure 7), as well as for a number
    of external early-type spirals (Figure 8).}
\item{The velocity dispersion both in the plane, $\sigma_r$, and perpendicular
    to it, $\sigma_z$ are similar to the observed velocity dispersions. Values obtained range from an average of around 6 km s$^{-1}$ 
    for $\epsilon$ = 0.05 to around 12 km s$^{-1}$ for $\epsilon$ = 0.2
    (Figure 5). For the smaller values of $\epsilon$ we find that $\sigma_r
    \approx \sigma_z$ , and note that values in the interarm regions
    are marginally smaller ($\approx$ 5 km s$^{-1}$) than in the arms
    ($\approx$ 7 km s$^{-1}$). For
    the higher values there is little arm/interarm dependence, but
    there is then a slight trend for $\sigma_z$ to be on average
    larger than $\sigma_r.$}
\end{enumerate}
\subsection{Molecular cloud properties}
\begin{enumerate}
\item{The mass spectra of clouds are of the form $dN/dM \propto
    M^{−1.9\pm0.1}$ for values of $\epsilon$ in the range $0.05 \leq
    \epsilon \leq 0.2$, with the masses being shifted to lower 
values for the higher values of $\epsilon$ (Figure 20).}
\item{The range of values of the $\alpha$ parameter which describes the
    degree to which the clouds are bound (see also
    \citealt{Dobbs2011}). For a surface density of $\Sigma = 8M_{\odot}$
    pc$^{-2}$, and for values of $\epsilon$ in the range
    $\epsilon=$ 0.05 -- 0.2 it is found that most clouds are unbound,
    or marginally bound,
    and have $\alpha$ in the range $\alpha \approx$ 1.0 -- 10. For smaller
    values of $\epsilon$ the models produce clouds
    which are disproportionately massive and strongly gravitationally
    bound. Higher surface density discs also tend to show more bound clouds
  for equivalent values of $\epsilon$.}
\item{ If molecular clouds formed simply by gravitational collapse
    from the ISM then conservation of angular momentum acquired from
    the galactic shear would ensure that they all rotate in the
    prograde direction. In fact, both in the Galaxy and in M33 about
    40 per cent of clouds rotate in the retrograde direction. As we
    have remarked before \citep{Dobbs2008}, if clouds build
    predominantly by collisional build-up of dense structures within
    an already inhomogeneous ISM, then the tendency of clouds to
    display both prograde and retrograde rotations in almost equal
    measure can be explained. We confirm this result in the current
    simulations (Figures 21 and 22). In Figure 22 we note that for the
    models with no spiral structure, in which the clouds build
    preferentially by self-gravity rather than collisions, the
    fraction of retrograde clouds is significantly reduced.}
\end{enumerate}

\subsection{Star formation}
\begin{enumerate}
\item{After initial transient behaviour, we find for $\epsilon
    \approx$ 0.05--0.2 that global star formation rates settle down to
    equilibrium values (Figures 11, 13 and 17) which depend on the
    average galactic gas surface density in agreement with the
    observed findings of Kennicutt (2008) (Figure 18). This result is
    independent of the presence or absence of an imposed spiral
    potential (cf. \citealt{Dobbs2009}). For efficiencies of
    $\epsilon \lesssim 0.5$ we find that the spiral potential gives rise to long-lived, massive, bound clouds in which star formation continues to accelerate over a long period; for these calculations we are unable to determine an equlibrium star formation rate.}
\item{We find that for an imposed spiral potential the distance from
    the galactic plane where star formation occurs depends
    significantly on the assumed value of $\epsilon$ (Figure 9). For the lower
    value of $\epsilon=0.05$ the star formation events occur with a
    scaleheight of 40 pc and do not stray more than around 150 pc
    above and below the galactic plane. In contrast, for $\epsilon=0.2$ the
    scaleheight of star formation events is around 80 pc, and events
    can occur up to around 500 pc from the plane.}
\end{enumerate}

In addition to these properties, we also obtain notable disagreements
with the observations when there is little or no feedback, as
exemplified by our calculation where $\epsilon=$ 0.01. In this
case, the mass spectrum is bimodal, there are very few retrograde
clouds, whilst the scale height of the ISM is too low in the absence
of feedback.

\section{Discussion}
We have used an imposed potential to simulate galaxies with strong 
spiral structure resulting from tidal interactions or central bars in
which the ISM is subject to strong shocks as it orbits in the galaxy. In
such galaxies, the ISM flows through the azimuthal inhomogeneities in 
the galactic potential (spiral arms) and so we might expect molecular
clouds to form mainly by collisional processes. We have not however
examined galaxies where self-gravity of the stellar component provides the dominant 
azimuthal component of the potential -- so called flocculent galaxies 
\citep{Li2005b,Gittins2006,DB2008,Fujii2011}. In this case the inhomogeneities 
occur close to co-rotation and so the relative velocity of the gas in the
spiral perturbations is strongly reduced. In such a scenario, we would
not necessarily expect to find the same properties of GMCs. For
example, we may well expect more prograde clouds compared to tidally
(or bar) driven spirals. In such galaxies, the clouds are likely to form
and disperse on timescales similar to the arm formation. A further
question is whether the formation of massive GMCs has any influence on
the stellar perturbations in the disc. We will address these issues in
future calculations which consistently model flocculent spiral
structure (see also \citealt{Wada2011}).

The main other work which has investigated cloud properties is
\citet{Tasker2011} and \citet{Tasker2009}. One main difference between
those papers and the current work is that \citet{Tasker2011} finds the
majority of molecular clouds have $\alpha<1$. \citet{Tasker2011} however
does not include supernovae feedback (nor a spiral potential), which by comparing Runs L1 and
L5 for example, is necessary to produce a population of unbound clouds. Instead
\citet{Tasker2011} adopts a temperature minimum, which presumably
prevents runaway gravitational collapse. \citet{Tasker2011} also finds
slightly lower fractions of retrograde rotating clouds. This however
is in agreement with our predictions that there are fewer retrograde
clouds in the absence of a spiral potential, and that the fraction of
retrograde clouds decreases with a less clumpy medium
(\citealt{Tasker2011} compares calculations with and without a diffuse
heating term). Both sets of calculations (see also \citealt{Dobbs2011}) 
find that clouds are relatively short-lived, i.e. $\lesssim 20$ Myr.

In our calculations, we observe the formation of large, massive clouds
which are not readily dispersed. They occur when a low amount of
energy is deposited as feedback, and are more frequent for discs with
higher surface densities. Such clouds would be likely to form massive 
star clusters. Though they studied radiation pressure, rather than the 
implicit momentum and thermal feedback we include here, 
\citet{Krumholz2010} also found that below a certain star formation 
efficiency (per free fall time), clouds were not disrupted. 
\citet{Swinbank2010} compare clouds in the Local Group, the nearby 
ULIRG Arp220 and the z=2 sub-millimetre galaxy SMMJ2135-0102.
 Interestingly they find that the clouds in the interacting galaxy
 (Arp220) and the z=2 galaxy are much more luminous for their 
physical size compared to local Group counterparts. The massive 
clouds in our models also display higher, and prolonged amounts of 
star formation. Although from our
  calculations, the formation of such clouds is found to be dependent
  on the efficiency, surface density and presence of spiral struture,
  we caution that our models may be too
  simplified (e.g. no magnetic fields, no
  time-dependent stellar feedback), whilst higher density calculations
  are not so well resolved, to make firm predictions.  Future work will involve models with more detailed 
stellar feedback, possibly in individual clouds, as well as more
realistic models of galaxies where massive clouds are common, such 
as starburst and high redshift galaxies. Such calculations may 
also indicate differences in the star formation efficiency, as
measured by the total mass of a given cloud converted into stars, 
and the Schmidt Kennicutt relation, in different environments.

\section{Acknowledgments}
The calculations presented in this paper were primarily performed on
the HLRB-II: SGI Altix 4700 supercomputer and the Linux cluster at the Leibniz supercomputer
centre, Garching. 
Many of the images were produced using \textsc{SPLASH}
\citep{splash2007}, a visualization tool for SPH that is publicly
available at http://www.astro.ex.ac.uk/people/dprice/splash. CLD
thanks Ian Bonnell for helpful discussions. We also thank an anonymous
referee for suggestions which improved the paper.

\appendix
\section{Supernovae feedback}
We add feedback according to the snowplough solution for a Sedov blast
wave test. The radius, velocity and temperature of the supernova bubble at a
time $t$ (in years) are

\begin{equation}
R_s=1.13 E_{51}^{115/511} n_0^{-135/511} t^{2/7}
\end{equation}
where $E_{51}$ is the supernovae energy in units of $10^{51}$ ergs and
$n_0$ is the ambient density in units of 1 cm$^{-3}$,
\begin{equation}
T_s=2.82\times 10^8 E_{51}^{134/511} n_0^{-24/511} t^{-4/7}
\end{equation}
and
\begin{equation}
V_s=413 n_0^{1/7} \zeta_m^{3/4} E_{51}^{1/14} (4/3 t_*-1/3)^{-7/10},
\end{equation}
 where 
\begin{equation}
t_*=t/t_{PDS}; \;  t_{PDS}=\frac{3.61 \times 10^4}{e} \frac{E_{51}^{3/14}}{\zeta_m^{5/14}n_0^{4/7}}  yr,
\end{equation}
$e$ is the natural logarithm and $\zeta_m$ is a metallicity factor set to
unity. These equations are taken from \citep{Ikeuchi1984} ($R_s$ and
$T_s$) and \citep{Cioffi1988} ($V_s$). The expressions from
\citet{Ikeuchi1984} utilise analytic calculations for blast waves from
a number of sources
\citep{Sedov1959,Woltjer1972,Chevalier1974,MCowie1977,McKee1977}. 
 
Our calculations provide $E_{51}$ (from Eqn.~1),
$R_{s}$, the radius of the region of gas and $n_0$, the ambient
density (calculated from a 50 pc width ring of gas surrounding the
region where feedback is inserted). We therefore rearrange Eqn.~A1 to
find $t$, which we then insert into equations A2 and A3 to find the
temperature and velocity. The values of $t$ are typically $10^4-10^5$
years and therefore slightly higher than the time at which the transition from the Sedov
to snowplough solution occurs.

We initially performed simulations of a single energy deposit in a 1
kpc cubed box. We found good agreement with the analytic results
provided the SPH timesteps were not too large and
the energy is not deposited in a tiny number of
particles. Interparticle penetration becomes problematic for long
timesteps (see also \citealt{Saitoh2008}), but we found this only
occurred with timesteps of $10^6$ years. In our global simulations,
the largest global timestep is much smaller than this. 
We also chose the radii of
the regions to insert supernovae so that there are typically a few 10's
of particles within those regions. 

\section{Numerical tests}
Using our adopted implementation of feedback, we do not add star
or sink particles, rather we simply insert energy above a certain
density criterion. With a mass resolution of 2500 or 5000 M$_{\odot}$,
we do not resolve the Jeans length at this criterion (though we note
that i) the densities of the particles in the molecular clouds are
typically one or two orders of magnitude below our feedback threshold,
and ii) the velocity dispersions are significantly higher than the
thermal sound speeds).
We did however test the dependence of our results on the criteria
chosen, as well as adopting a temperature floor of 500 K.
 
In Fig.~B1 we show the structure of the disc with density crtieria of
100, $10^4$ cm$^{-3}$, and a temperature floor of 500 K as well as a
low resolution calculation with 250,000 particles\footnote{Typically
  gas at density thresholds much lower than 100 cm$^{-3}$ does not meet the requirements for
  inserting feedback in our prescription, i.e. boundedness, a converging flow.}. The structure of
the disc is similar with the different density criteria. Though with
the low density criterion, there is less molecular gas so there is
slightly less feedback, and more bound clouds, giving an evolution
more similar to Run L1. The global structure
of the disc is also similar with the temperature floor, and the lower
resolution, but the small scale structure is not captured.

We also checked the cloud properties in these (1 million particle)
calculations. 
The clouds in the model with the very high density criterion are very
similar compared to the clouds in L5. The mass spectrum (though with a slightly
lower normalisation) has the same slope, and the distribution of
angular momenta are very similar. For the model with the lower
threshold, there are more bound clouds, and a smaller fraction of
retrograde clouds. However this is to be expected given
that the efficiency is effectively less in this calculation. For the
calculation with a temperature floor of 500 K, the mass spectrum is
shallower, probably since there is less small scale structure. There
are also fewer (27 per cent) retrograde clouds, again presumably
because the gas is less clumpy.
\begin{figure*}
\centerline{
\includegraphics[scale=0.52]{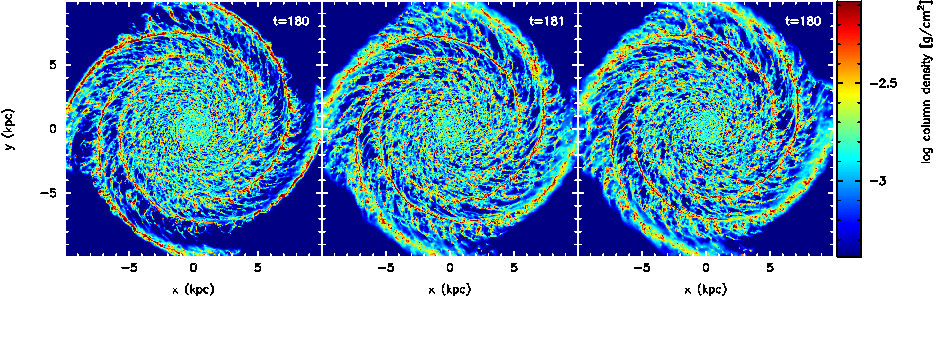}}
\centerline{
\includegraphics[scale=0.24]{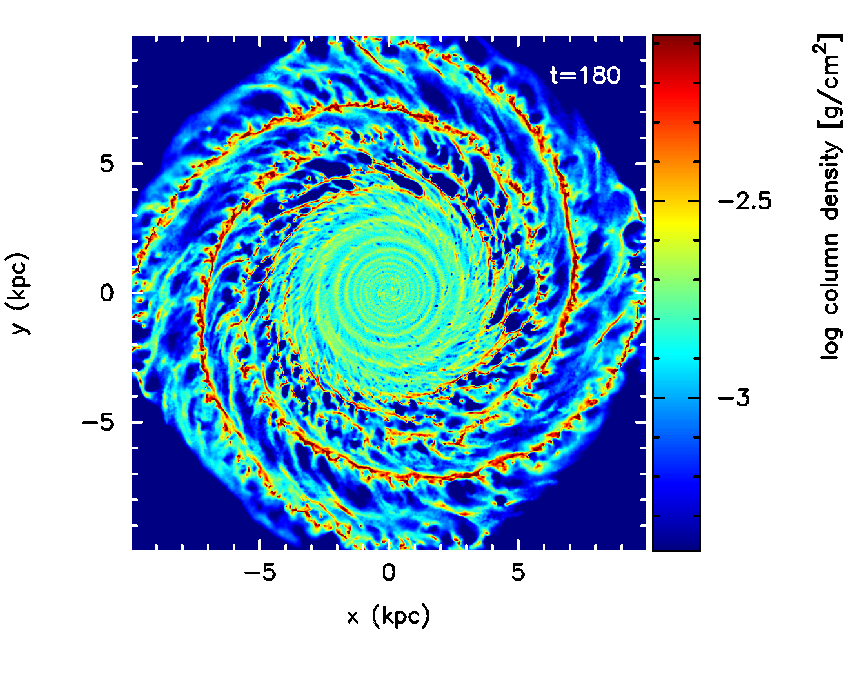}
\includegraphics[scale=0.24]{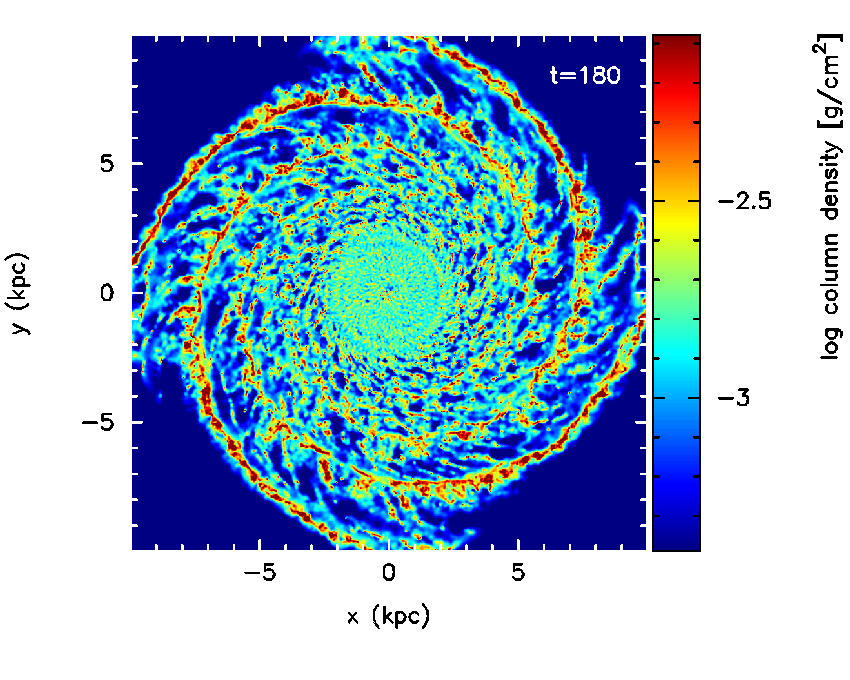}}
\caption{The column density is shown for calculations with a feedback density
  threshold of 100, 1000 and $10^4$ cm$^{-3}$ (top row), where a
  temperature floor of 500 K is imposed (lower left) and with a
  resolution of 250,000 particles (lower right). The large scale
  structure is similar in all cases, though the small scale structure
  is not captured with the imposed temperature floor, or at low
  resolution. The structure for the case with a threshold of 100
  cm$^{-3}$ is more similar to Run L1, since the efficiency of star
  formation is effectively lower.}
\end{figure*}
\bibliographystyle{mn2e}
\bibliography{Dobbs}
\bsp
\label{lastpage}
\end{document}